\documentclass[10pt]{aastex}
\usepackage{emulateapj5}
\usepackage{apjfonts}
\usepackage{psfig}

\newcommand{\kms}       {\mbox{km s$^{-1}$}}%
\newcommand{\kmsMpc}	{\mbox{km s$^{-1}$ Mpc$^{-1}$}}%

\shorttitle{Spurious Luminosity Offsets}
\shortauthors{Kannappan \& Barton}

\slugcomment{Accepted by AJ 2004 February 5}

\begin{document}

\title{
Tools for Identifying Spurious Luminosity Offsets in Tully-Fisher Studies: \\
Application at Low Redshift and Implications for High Redshift
}

\author{Sheila J. Kannappan\altaffilmark{1} \&
Elizabeth J. Barton\altaffilmark{2}}

\altaffiltext{1}{Harlan Smith Fellow, McDonald Observatory,
The University of Texas at Austin, 1 University Station C1402,
Austin, TX 78712-0259; sheila@astro.as.utexas.edu}

\altaffiltext{2}{Hubble Fellow, 
The University of Arizona, Steward Observatory, 933
N. Cherry Ave., Tucson, AZ 85721; ebarton@as.arizona.edu}

\begin{abstract}
Studies of high-redshift galaxies usually interpret offsets from the
Tully-Fisher (TF) relation as luminosity evolution.  However, apparent
luminosity offsets may actually reflect anomalous velocity widths.
Rotation curve anomalies such as strong asymmetries or radial
truncation are probably common in high-$z$ samples, due to frequent
galaxy interactions and in some cases low S/N data, although low
physical resolution may mask these anomalies.  In this paper we
analyze well-resolved, one-dimensional optical emission-line rotation
curves from two low-$z$ samples: the Close Pairs Survey, which
contains a high frequency of interacting galaxies, and the Nearby
Field Galaxy Survey (NFGS), which represents the general galaxy
population.  Unlike most low-$z$ TF samples, but in the spirit of many
high-$z$ samples, these surveys reflect the natural diversity of
emission-line galaxy morphologies, including peculiar, interacting,
and early-type galaxies.  We adopt objective, quantitative criteria to
reject galaxies with severe kinematic anomalies, and we use a
statistical velocity width measure that is insensitive to minor
kinematic distortions.  Severely anomalous galaxies are roughly twice
as frequent in the Close Pairs Survey as in the NFGS, and these
galaxies' TF offsets collectively resemble the ``differential
luminosity evolution'' claimed in some high-$z$ studies, with larger
offsets at lower luminosities.  With the anomalous galaxies rejected,
however, the TF relations for the Close Pairs Survey and the NFGS are
quite similar.  Furthermore, the two surveys follow very similar
relations between color and TF residuals.  The Close Pairs Survey
color--TF residual relation extends to bluer colors and brighter TF
residuals.  Strong outliers from this relation are virtually always
kinematically anomalous.  As a result, the color--TF residual relation
can serve as a powerful tool for separating reliable luminosity
offsets from offsets associated with kinematic anomalies.  This tool
may prove especially useful at high $z$, where direct detection of
kinematic distortions is not always feasible.  Although we cannot
reliably measure luminosity evolution for galaxies with kinematic
anomalies, the TF offsets associated with these anomalies may offer a
sensitive probe of evolution in the frequency and intensity of mergers
and interactions on different mass scales.  We perform a preliminary
reanalysis of high-$z$ TF data from the FORS Deep Field and find: (1)
overall luminosity evolution of $\sim$0.3 mag; (2) strong slope
evolution driven by kinematically anomalous galaxies, which show TF
offsets of up to $\sim$2 mag at low luminosities; and (3) an
additional zero-point offset of $\sim$0.2 mag linked to kinematically
anomalous galaxies.
\end{abstract}


\section{Introduction}
\label{sc:intro}

As the fundamental scaling relation between luminosity and rotation
velocity for disk galaxies, the Tully-Fisher relation \citep[TF
relation,][]{tully.fisher:new} evolves along with the galaxies that define
it, reflecting general trends in mass assembly and star formation.
Numerous studies have sought to trace the star formation history of disk
galaxies via the redshift evolution of the zero point of the TF relation,
under the assumption that TF zero point offsets represent luminosity
evolution.  Awkwardly, some studies find minimal luminosity evolution to
redshifts as high as $z\sim1$
\citep[e.g.,][]{forbes.phillips.ea:keck,vogt.phillips.ea:optical,bershady.haynes.ea:rotation},
while others report substantial 1.5--2 mag offsets at lower redshifts
\citep[e.g.,][]{rix.guhathakurta.ea:internal,simard.pritchet:internal}.
Efforts to reconcile these results have generally invoked differential
evolution, in which only low-luminosity galaxies evolve significantly
\citep[e.g.,][]{simard.pritchet:internal,ziegler.b-ohm.ea:evolution}.

However, some ``luminosity offsets'' may actually be velocity offsets.  In
the TF relation, underestimated rotation velocities look exactly like
enhanced luminosities.  The following three examples are particularly
relevant to high $z$ studies.

(1) Optical emission-line data for high-$z$ ($z$ $\sim$ 0.25--1 for this
paper) blue compact galaxies may not extend to large enough radii to sample
peak rotation velocities, based on 21 cm HI studies of analogous galaxies
at low $z$ (\citeauthor{barton.:possible} 2001;
\citeauthor{pisano.kobulnicky.ea:gas} 2001; see also
\citeauthor{kobulnicky.gebhardt:obtaining} 2000;
\citeauthor{courteau.sohn:galaxy} 2003).  Of course, high-$z$ studies must
employ optical lines rather than HI to measure rotation velocities.  Also,
most high-$z$ studies have selection biases favoring the bright galaxy
cores and strong emission lines typical of blue compact galaxies.  One
might hope that high-$z$ studies would be insensitive to radially truncated
emission-line data, because unlike low-$z$ analyses, high-$z$ analyses
usually derive rotation velocities by analyzing kinematic and photometric
profiles together
\citep[e.g.,][]{vogt.forbes.ea:optical,simard.pritchet:analysis,ziegler.b-ohm.ea:evolution}. However,
such modeling techniques typically rely on simplifying assumptions that
blue compact galaxies probably routinely violate, such as the assumption of
a close correspondence between emission-line and underlying disk-continuum
fluxes, or the assumption that rotation curves can be simply parametrised
based on exponential-disk fits to the spatial flux distribution regardless
of disturbances or central mass concentrations.

(2) Low
S/N can also cause radially truncated emission and underestimated
rotation velocities, especially in samples already biased toward
galaxies with centrally concentrated emission.  Accounting for
S/N-induced rotation curve truncation could significantly reduce
discrepancies between high-$z$ TF studies \citep[as discussed for the
Simard \& Pritchet and Vogt et al.\ studies
by][]{kannappan:kinematic}.

(3) Distorted rotation curves may also yield unreliable rotation
estimates and systematic velocity offsets.  In low-$z$ TF samples that
contain interacting or morphologically peculiar galaxies, disturbances
in longslit optical emission-line rotation curves clearly correlate
with apparent luminosity boosts from the TF relation
\citep[][]{barton.geller.ea:tully-fisher,kannappan.fabricant.ea:physical}.
We suspect that these apparent boosts may not be pure luminosity
offsets, especially when they are larger than would be expected based
on colors or H$\alpha$ equivalent widths
\citep{kannappan.fabricant.ea:physical}.

At low $z$, large TF offsets associated with distorted or truncated
rotation curves are most common for low-luminosity galaxies, and the
affected galaxies often display emission-line S0 or irregular
morphologies, sometimes with independent evidence of interactions
(\citeauthor{kannappan.fabricant.ea:physical} 2002; see also
\citeauthor{kobulnicky.gebhardt:obtaining} 2000;
\citeauthor{barton.geller.ea:tully-fisher} 2001).  Although most
low-$z$ TF studies would reject such galaxies
\citep[e.g.,][]{courteau:optical,haynes.giovanelli.ea:i-band,tully.pierce:distances},
all of the high-$z$ studies that report substantial faint-end
luminosity evolution employ selection criteria that would admit them
\citep[e.g.,][]{rix.guhathakurta.ea:internal,simard.pritchet:internal,mall-en-ornelas.lilly.ea:internal,ziegler.b-ohm.ea:evolution}.
Furthermore, the frequency of such galaxies may be enhanced in
high-$z$ samples to the extent that the interaction rate increases
with $z$ \citep{patton.pritchet.ea:dynamically,murali.katz.ea:growth}.
These points raise the obvious concern that high-$z$ TF samples may
contain a population of galaxies whose velocity offsets mimic
differential luminosity evolution.

Another key consideration in interpreting apparent luminosity offsets
is the possibility of third-parameter dependence in TF residuals.
Numerous studies have examined possible physical drivers of TF
offsets, including morphology, surface brightness, gas content,
environment, and color, for TF samples chosen by a variety of criteria
\citep[e.g.,][ and additional references
therein]{roberts:twenty-one,rubin.burstein.ea:rotation,giraud:two-color,pierce.tully:distances,mould.han.ea:nonlinearity,pierce.tully:luminosity-line,sprayberry.bernstein.ea:mass-to-light,courteau.rix:maximal,mcgaugh.schombert.ea:baryonic,verheijen:ursa*1,barton.geller.ea:tully-fisher,kannappan.fabricant.ea:physical}.
In a recent analysis, \citet{kannappan.fabricant.ea:physical}
demonstrate that TF residuals correlate strongly with star formation
indicators --- color and EW(H$\alpha$) --- in the Nearby Field Galaxy
Survey
\citep[NFGS,][]{jansen.franx.ea:surface,kannappan.fabricant.ea:physical},
a statistically representative survey of all galaxy types with no bias
against interacting, peculiar, or early-type galaxies.  The inclusion
of such galaxies distinguishes the NFGS TF sample (and most high-$z$
TF samples) from the majority of low-$z$ TF samples, which restrict
analysis to a limited range of morphologies that may show only weak
correlations between color and TF residuals \citep[e.g.,][ see
Kannappan et al.\ 2002 for further discussion]{courteau.rix:maximal}.
However, the Ursa Major cluster sample of
\citet{verheijen.sancisi:ursa}, which approximates a volume-limited
sample, shows a stronger color--TF residual correlation
\citep{verheijen:ursa*1,kannappan.fabricant.ea:physical}, and
\citet{bershady.haynes.ea:rotation} also find initial evidence for a
color--TF residual correlation at high $z$.  The existence of this
correlation implies that high-$z$ samples that differ in average color
because of different selection criteria will also differ in average TF
zero-point offset.  If high-$z$ galaxies follow the same color--TF
residual relation the NFGS follows, then we can use this relation to
correct high-$z$ TF offsets for any bias toward blue colors (or we can
use the EW(H$\alpha$)--TF residual relation to correct for any bias
toward strong emission lines).  Moreover, once such biases are
removed, we can compare the remaining zero-point offset with the
luminosity evolution predicted by the color--TF residual relation
(based on true differences in mean color between high and low $z$), in
order to determine whether TF zero-point evolution includes not only
luminosity evolution, but also additional evolution reflecting the
growth of stellar-to-total mass fractions over cosmic time
\citep{kannappan.gillespie.ea:interpreting}.

Obtaining a well-defined color--TF residual (CTFR) relation and
measuring evolutionary offsets reliably may require special attention
to galaxies with distorted or radially truncated rotation curves.  In
this paper, we demonstrate such an analysis at low $z$ using the Close
Pairs Survey of \citet{barton.geller.ea:tully-fisher}.  The
interacting galaxies in this survey display luminosity enhancements
and misleading velocity offsets much like high-$z$ galaxies, as
previously shown by \citet{barton.geller.ea:tully-fisher}.  However,
at low $z$ we can use high-resolution kinematic data to identify
problem rotation curves objectively, using quantitative tests of
radial truncation and asymmetry of shape based on those introduced by
\citet{kannappan.fabricant.ea:physical}.  Without explicitly
accounting for kinematic anomalies,
\citeauthor{barton.geller.ea:tully-fisher} could not decouple
luminosity offsets from velocity offsets and found no statistically
significant CTFR relation for the Close Pairs Survey.  We recover the
CTFR relation for the Close Pairs Survey by eliminating galaxies with
severely truncated or asymmetric rotation curves based on quantitative
criteria, and by analyzing modestly asymmetric rotation curves with a
robust velocity width measure that does not assume a functional form.
Using these procedures, we find that the TF and CTFR relations
for the Close Pairs Survey look very similar to the corresponding relations
for the NFGS.  Furthermore, the tightness of the Close Pairs Survey
CTFR relation suggests that if a similar relation holds at higher $z$,
determining whether galaxies lie on or off its locus may serve as a
way to distinguish reliable luminosity evolution from TF
offsets associated with kinematic anomalies.

Below, we describe the Close Pairs Survey and the NFGS (\S~2), as well as
our analysis methods (\S~3), including quantitative criteria for
identifying strongly asymmetric or truncated rotation curves.  We then
analyze the TF and CTFR relations for the Close Pairs Survey, with
attention to kinematic anomalies, and compare the Close Pairs Survey
relations to the corresponding NFGS relations (\S~4).  In \S~5 we examine
the possible drivers of kinematic anomalies. We go on to consider the
implications of our results for high-$z$ TF studies in \S~6.  Finally, we
summarize our conclusions in \S~7.

\section{Data}
\label{sc:data}

Our analysis makes use of two statistical surveys drawn from the CfA
redshift surveys \citep[][]{geller.huchra:mapping}: the Nearby Field
Galaxy Survey
\citep[][]{jansen.fabricant.ea:spectrophotometry,jansen.franx.ea:surface,kannappan.fabricant.ea:physical},
representing the general galaxy population, and the Close Pairs Survey
\citep[][]{barton.geller.ea:tidally,barton.geller.ea:tully-fisher},
representing galaxy pairs with line-of-sight velocity separation
$\Delta V < 1000$ \kms\ and projected spatial separation $\Delta X \la
100$ kpc (the distance limit differs from the original reference
because we quote all distances and magnitudes using H$_0=75$ \kmsMpc\
and correct both surveys for Virgocentric infall following
\citeauthor[][]{jansen.franx.ea:surface} 2000 and
\citeauthor{kraan-korteweg.sandage.ea:effect} 1984).  These two
surveys were selected without explicit bias in morphology or global
environment.  The Close Pairs Survey reflects the inherent luminosity
bias of its magnitude-limited parent survey and also explicitly
excludes galaxies with redshifts below 2300 \kms.  In contrast, the
NFGS was selected with a greater representation of low-luminosity
galaxies, in an effort to reproduce the local galaxy luminosity
function \citep{jansen.franx.ea:surface}; in practice, the NFGS
luminosity distribution varies slowly over the range $-16>M_{B}>-22$
and cuts off for brighter and fainter galaxies, with emission-line
galaxies naturally favoring lower luminosities within the sample
\citep[see][]{kannappan.fabricant.ea:physical}.  Both surveys transmit
the surface-brightness bias of their parent surveys, although the NFGS
selection procedure was designed to minimize this bias
\citep{jansen.franx.ea:surface}.

For TF analysis we use one-dimensional optical emission-line rotation
curves obtained with the FAST spectrograph on the 60-inch Tillinghast
telescope at Mt. Hopkins (NFGS) and the Blue Channel spectrograph on the
pre-conversion MMT (Close Pairs Survey and a few NFGS galaxies), as
described in \citet{barton.geller.ea:tully-fisher} and
\citet{kannappan.fabricant.ea:physical}.  Besides considering only
emission-line galaxies, we further restrict our analysis to galaxies with
$i>40$ and M$_{\rm B}^i < -18$, except for a schematic look at TF outlier
behavior among NFGS dwarfs in \S~\ref{sc:hizsec}.  We also require good
alignment between the spectrograph slit and the galaxy major axis
($\Delta$P.A. $<$ 10 for the NFGS and $\Delta$P.A. $<$ 20 for the Close
Pairs Survey\footnote{The different misalignment requirements for the two
surveys have no effect on their relative frequencies of kinematic
anomalies, as misalignments $\la$20$\degr$ are too small to cause the
severe anomalies considered in this paper.  The eight Close Pairs Survey
galaxies with 10 $<$ $\Delta$P.A. $<$ 20 have rates of rotation curve
asymmetry and rotation curve truncation consistent with overall rates for
the survey as a whole.}).

Fig.~\ref{fg:twosampprops} compares property distributions for the two TF
samples, in luminosity, morphology, color, redshift, surface brightness,
and global environmental density.  Sample properties are broadly similar,
except for the luminosity and redshift distributions, which reflect the
different selection methods discussed above.  Notably, the choice of {\em
local} pair environments does not strongly affect the distribution of {\em
global} density environments for Close Pairs Survey galaxies, though these
galaxies may show a slight underrepresentation of the lowest density
environments compared to the NFGS (Fig.~\ref{fg:twosampprops}f).  The
requirement of detectable emission lines implies that within the NFGS and
the Close Pairs Survey, the subsamples used for TF analysis have a higher
proportion of low-density environments, as many cluster galaxies lack
significant emission.  Most high-$z$ TF samples also emphasize field
environments, for similar reasons.

The NFGS and the Close Pairs Survey have three TF galaxies in common
(A00442+3224, A22551+1931N, and NGC~7537).  For these three galaxies, raw
velocity widths from optical rotation curves are in excellent agreement,
all within $\sim$8 \kms.  We also find reasonable agreement in rotation
curve structure, despite different rotation curve extraction techniques
(\S~\ref{sc:repro}). Redshifts agree within 15--50 \kms, and effective
colors agree within 0.02 mag.  Total B-band magnitudes agree within 0.15
mag.  The only parameter for which the two surveys do not track closely is
inclination angle (or equivalently, axial ratio), where we find differences
of 10--20$\degr$ in both directions for the three common galaxies.  Close
Pairs Survey inclinations derive from careful analysis of new CCD data,
whereas NFGS inclinations derive from low-precision diameter measurements
tabulated in the UGC \citep{nilson:uppsala}.

\vspace{1.0in}

\section{Methods}

\subsection{Tully-Fisher Analysis Techniques}
\label{sc:fittech}

To facilitate direct comparison of the two surveys, we recompute the
velocity widths and inclinations for the Close Pairs Survey \citep[][
hereafter B01]{barton.geller.ea:tully-fisher} with the methods used for the
NFGS \citep[see][ hereafter K02]{kannappan.fabricant.ea:physical}, and we
correct the Close Pairs redshifts for Virgocentric infall to match the NFGS
(\S~\ref{sc:data}).  We calculate extinction corrections for both surveys
using the results of \citet{tully.pierce.ea:global} as described in K02,
but without K02's special treatment of S0 galaxies.

The new inclinations, extinction corrections, and redshifts have
quantitative, but no qualitative effects on the results.  However, the new
velocity widths do lead to some qualitative differences, because the
``probable min-max'' velocity measure $V_{pmm}$ adopted by K02 following
\citet{raychaudhury..ea:tests} is more robust than the velocity measure
$V_{2.2}$ adopted by B01 following \citet{courteau:optical}.  $V_{2.2}$
requires that a rotation curve conform to a standard functional form, which
may provide a poor fit to rotation curves distorted by interaction. 
In contrast, $V_{pmm}$ uses all of the data points without imposing a
particular model.  As in K02, we define $V_{pmm} = 0.5(V_{pmax}-V_{pmin})$,
where $V_{pmax/pmin}$ is defined as having a 10\% chance of exceeding/lying
below all velocities in the rotation curve.  Each data point is
modeled as a Gaussian distribution about the measured value, with $\sigma$
equal to the measurement error (see K02 for formulae).  Using $V_{pmm}$,
the pattern of TF offsets changes in such a way as to clarify the
correlations reported in \S~\ref{sc:pairsresults}.  (The parameter
$W_{V_{pmm}}^i$ is related to $V_{pmm}$ via an inclination correction and a
linear transformation that puts it on the same scale as the $W_{50}$
linewidths of radio observers, see K02.)

B01 and K02 have discussed the pros and cons of various TF fitting
techniques.  To avoid slope bias, we adopt an unweighted ``inverse''
fit (minimizing residuals in velocity) as our primary technique.  A
bias-corrected ``forward'' fit \citep{willick:statistical} would give
similar results, but such a fit would be very difficult to implement
for the NFGS because of the intricacy of the survey's statistical
selection procedure \citep{jansen.franx.ea:surface}.  In
\S~\ref{sc:pairsresults} we consider the effect of using a
bias-corrected forward fit for the Close Pairs Survey, following the
methods of B01, and we find no significant change in the results.
Note that kinematically anomalous galaxies (\S~\ref{sc:quant}) are
excluded from all TF fits, although they appear with special symbols
in TF plots.

Ideally, we would like to compute the TF residuals used for the CTFR
relation according to the procedure described in K02, in which the
reference TF relation is defined by a fit over intermediate luminosities
($-17>{\rm M}_{\rm B}^{i}>-21$) to avoid bias from high- and low-luminosity
galaxies whose TF scatter is asymmetric.  Unfortunately, this procedure is
not practical for the Close Pairs Survey, both because the survey was not
intended as a complete or representative sample of TF galaxies, and because
the survey includes few galaxies fainter than $\sim-19.5$.  In fact, the
low-luminosity cutoff we apply to the NFGS at $-18$ is not necessary for
the Close Pairs Survey, which includes no galaxies fainter than $-18$
because of its low-redshift cutoff (\S~\ref{sc:data}).  We therefore
analyze TF residuals for the Close Pairs Survey in two ways: (1) purely
internally, i.e.\ determining both the TF relation and the CTFR relation
from the Close Pairs Survey itself, with no luminosity cuts
(\S~\ref{sc:pairsresults}); and (2) using the NFGS as a reference, i.e.\
determining the CTFR relation using Close Pairs TF residuals measured
relative to the NFGS TF relation, again with no upper luminosity cuts (but
with the NFGS limited to $M_{B}^{i}<-18$) to simplify comparison
(\S~\ref{sc:combresults}).

\subsection{Identifying Kinematically Anomalous Galaxies}
\label{sc:quant}

We refer to galaxies whose rotation curves are severely truncated in
extent or highly asymmetric in shape as kinematically anomalous.
Fig.~\ref{fg:rcegs} shows several examples of both anomalous and
normal rotation curves.  The next two sections discuss the criteria we
use to flag kinematic anomalies. Both B01 and K02 have described
methods for identifying anomalous rotation curves; here we adapt the
methods of K02, who define continuous, quantitative measures of
truncation and asymmetry. Although these measures are objective, the
cutoff values of asymmetry and truncation used to reject galaxies must
be empirically determined in order to optimize the rejection of TF and
CTFR outliers for a given sample.  Furthermore, strong asymmetries and
severe truncation rarely coexist (\S~\ref{sc:asym} and \ref{sc:interp}),
so both types of anomaly must be considered in order to reject
outliers successfully.  We note that our empirical approach does not
depend on whether the observed anomalies are intrinsic to the target
galaxies or just artifacts of the data; however, we will argue in
\S\ref{sc:interp} that at least some anomalies are intrinsic and
discuss their physical origin.

\subsubsection{Rotation Curve Truncation}
\label{sc:trunc}

To evaluate rotation curve truncation, we consider the average of the radial
extents on the two sides of the rotation curve.  (This average measure is slightly
more robust than the one-sided measure used by K02.)  For a pure
theoretical exponential disk, the rotation curve will reach maximum velocity at
2.2 disk scale lengths or $\sim$1.3$r_e$
\citep[]{freeman:on}, so a rotation curve extent of less than
1.3$r_e$ could be considered suspect.  In practice however, galaxies
with rotation curves extending to $\sim$1.0--1.3 $r_e$ are not outliers in our TF
and CTFR relations, so we flag galaxies as anomalous only if their rotation curves
are truncated at $<$0.9$r_e$.  Because measurements of $r_e$ are
sensitive to details of profile extrapolation, the exact cutoff used
to identify truncated rotation curves should be determined within a given TF data
set.

\subsubsection{Rotation Curve Asymmetry}
\label{sc:asym}

Following K02, we define rotation curve asymmetry as the mean absolute
deviation between velocities on the two sides of the rotation curve,
expressed as a percentage of the velocity width 2$V_{pmm}$.  This
definition quantifies asymmetries in velocity structure between the
two sides of the curve, e.g., due to one side rising and the other
falling.  We measure asymmetries by a procedure that involves
numerically searching for the coordinate center of the rotation curve
that minimizes the inner asymmetry (inside 1.3$r_e$) within certain
constraints (for full details see K02).  A key constraint is that the
spatial center must stay within the error bars of the continuum peak
position.\footnote{We have experimented with varying the error
constraint, and we find that the results do not change as long as the
spatial center is required to stay within $\pm0.5$ pixel of the
initial continuum peak position estimate (equal to $\pm0\farcs6$ for
our preferred binning of the Close Pairs Survey data).  Asymmetries
shown in this paper were computed with stricter constraints, typically
$\pm0.15$ pixel (equal to $\pm0\farcs2$ for our preferred binning of
the Close Pairs Survey data). For galaxies with two continuum peaks,
we have attempted to choose the most appropriate continuum peak by
eye, but in two cases, we switched to the second continuum peak after
seeing that the first continuum peak was further from the kinematic
center.}  Therefore galaxies in which the continuum center and the gas
kinematic center do not agree tend to have large asymmetries.  In
essence, the asymmetry index combines a measure of shape asymmetry
with a measure of the offset between the center of stellar light and
the center of gas motion, due to dynamical disequilibrium or possibly
extreme dust extinction.  Note that the choice of center has no effect
on $V_{pmm}$ or on our measure of rotation curve truncation, which
averages the spatial extent on the two sides of the rotation curve.

We adopt a purely empirical definition of ``strong'' rotation curve
asymmetry, based on the observed asymmetry distributions for both
surveys (Fig.~\ref{fg:asymdist}a).  Most galaxies have asymmetries
$\la$8\%, but a few form a higher asymmetry tail to the distribution.
These galaxies also scatter outside the rotation curve
asymmetry--luminosity correlation reported by K02
(Fig.~\ref{fg:asymdist}b).  However, we find that most galaxies with
moderately strong rotation curve asymmetries (8-10\%) and no rotation
curve truncation follow the TF and CTFR relations, so we flag galaxies
as anomalous only for rotation curve asymmetries $>$10\%.

Notably, strong asymmetries are rare among the most truncated rotation
curves, although there is no clear correlation between asymmetry and
rotation curve extent (Fig.~\ref{fg:truncasym}).  The pattern in
Fig.~\ref{fg:truncasym} is consistent with the view that severe truncation
and strong asymmetries arise from related physical causes
(\S~\ref{sc:interp}), but the most extreme truncation leads to a loss of
information, where rotation curves may simply have inadequate radial extent
to reflect asymmetries that would otherwise be significant.

Measuring rotation curve asymmetries reliably also requires adequate
spatial resolution.  Fig.~\ref{fg:binning} illustrates the effect of
degrading resolution for the Close Pairs Survey.  We show rotation curve
asymmetries determined with the original spatial sampling of 0.6'' per
pixel (1--2'' seeing) and with the data binned by 2 and by 4.  At the
lowest resolution, some information is lost and occasionally the code
crashes with insufficient data.  We adopt the binned by 2 results for this
paper because higher resolution is not available for two of the Close Pairs
Survey galaxies, and because there are only minor differences between the
binned by 2 and binned by 1 results.

For the NFGS, most rotation curves were binned on readout to 2.27''
per pixel (2'' seeing).  While this resolution is lower than the
resolution used for the Close Pairs Survey, the loss of information is
mostly offset by the fact that the NFGS TF sample is $\sim$1.7 times
closer in median redshift than the Close Pairs TF sample
(Fig.~\ref{fg:twosampprops}d).

\subsubsection{Reproducibility of Truncation and Asymmetry Measures}
\label{sc:repro}

We evaluate the reproducibility of our truncation and asymmetry
measures by comparing results for the three galaxies common to the
NFGS and the Close Pairs Survey (observed with the FAST/60-inch
combination for the NFGS and the Blue Channel/pre-conversion MMT
combination for the Close Pairs Survey).  Raw rotation curves from the
two data sets agree well (Fig.~\ref{fg:threecmp}).  In two cases we
see small deviations in the inner rise region that probably reflect
differences in spatial resolution and rotation curve extraction
technique between the surveys.\footnote{\label{fn:struct} For
A22551+1931N, whose rotation curve extends to only $\pm$5 arcsec, the
two times lower resolution of the NFGS data may partly account for the
shallower rise of the NFGS rotation curve compared to the Close Pairs
Survey rotation curve.  In addition, for both A00442+3224 and
A22551+1931N, small differences in rotation curve structure may arise
because these galaxies have complex line profiles in their central
regions (A00442+3224 shows multiple components and A22551+1931N shows
asymmetric wings).  In such cases, we expect discrepancies between the
two surveys, because NFGS rotation curves were extracted from 2D CCD
spectra using simultaneous Gaussian fits to multiple emission lines,
while Close Pairs Survey rotation curves were extracted using cross
correlation.  As discussed in \citet{barton.kannappan.ea:rotation},
these two techniques respond differently to non-Gaussian or
multiple-component line profiles: cross correlation seeks out the
peak, while Gaussian fitting finds something closer to an
emission-weighted average.} Despite these small deviations, asymmetry
measurements show the same general pattern in both data sets.

Truncation results also agree, in the sense that none of the rotation
curves in either data set falls short of the cutoff radius at 0.9$r_e$
(shown in Fig.~\ref{fg:threecmp} with dashed and dotted gray lines for the
Close Pairs Survey and NFGS respectively).  However, we see up to 30\%
disagreement in measured $r_e$, presumably due to different techniques of
photometric profile extrapolation.  Rotation curve extents also differ,
with NFGS rotation curves generally extending further.  Examination of the
raw data indicates lower S/N in the Close Pairs data.  The individual
points in the NFGS rotation curves would have to have $\sim$4$\times$ lower
S/N to yield similar rotation curve extents for the two surveys (with the
NFGS rejection threshold set at S/N = 3 and the Close Pairs rejection
threshold set at cross-correlation R = 2).\footnote{We have tested more
complicated rejection algorithms for individual points in the Close Pairs
rotation curves, with minimal effect on rotation curve extents.  In
particular, truncated rotation curves remain truncated.}  These survey
differences confirm that estimates of rotation curve truncation are subject
to noise and systematic effects.  We therefore reiterate that the exact
rotation curve truncation threshold used for rejection should be determined
within a given TF data set.

\section{Tully-Fisher Results}

This section discusses TF results obtained by excluding galaxies with
severely truncated or asymmetric rotation curves.  We defer interpretation
of these kinematic anomalies to \S~\ref{sc:interp}.

\subsection{The Close Pairs Survey}
\label{sc:pairsresults}

Because the previous analysis of B01 revealed no significant
correlation between color and TF residuals for the Close Pairs Survey,
we must first demonstrate that our claim of a CTFR relation in the
Close Pairs Survey is robust.  Below we show how our treatment of
kinematic anomalies allows us to detect the CTFR relation, in
particular because of the very close correspondence between
kinematically anomalous galaxies and CTFR outliers (which constitute
$\sim$10\% of the Close Pairs Survey TF sample).

Fig.~\ref{fg:pairsalone} shows the TF and CTFR relations for the Close
Pairs Survey, analyzing the survey purely on its own.  The TF and CTFR fits
exclude galaxies with highly truncated or asymmetric rotation curves
(extent $<$ 0.9$r_e$ or asymmetry $>$ 10\%, \S~\ref{sc:quant}), which are
marked with triangles and circles respectively.  Without these
kinematically anomalous galaxies, the CTFR relation emerges clearly.
Defining TF residuals relative to the inverse-fit TF relation shown by the
solid line, a Spearman rank test gives 6.5$\times10^{-11}$ probability of
no correlation.  Using a bias-corrected forward-fit TF relation based on
the methods of B01 also yields a strong CTFR relation (no-correlation
probability 1.3$\times10^{-7}$), because the forward-fit and inverse-fit TF
slopes are very similar (solid and dotted lines in
Fig.~\ref{fg:pairsalone}, both shown with the zero point from the inverse
fit.\footnote{Bias-correcting TF fits generally yield zero point shifts
that vary with the model assumed for the luminosity and color dependence of
TF scatter, as well as with the details of how measurement errors influence
sample selection. This zero point shift has no relevance to our analysis,
so we choose not to model it.})

As discussed by B01 and K02, TF residual correlations like the CTFR
relation should be tested rigorously, since any parameter that depends on
luminosity will correlate with TF residuals if the TF slope is incorrect.
K02 describe a fitting algorithm that avoids this problem, but this
algorithm is not ideal for a sample like the Close Pairs Survey with a
top-heavy luminosity distribution (\S~\ref{sc:fittech}). Therefore we adopt
the strategy of B01, using the robustness of the CTFR relation under
changes of slope as a sanity check.  The dashed line in
Fig.~\ref{fg:pairsalone} shows the slope required to eliminate the CTFR
relation (i.e., to increase the probability of no correlation to 10\% in a
Spearman rank test).  This slope is implausibly shallow and does not
even pass through the bright end of the TF relation.  Even the fit
including kinematically anomalous galaxies (gray dot-dashed line) is
significantly steeper, confirming the reality of the CTFR relation.

Compared to sigma-clipping applied to the TF relation (e.g., B01), our
technique of identifying anomalous galaxies from their rotation curve properties is
very effective at isolating galaxies whose TF residuals do not follow the
CTFR relation.  Most kinematically anomalous galaxies are CTFR outliers.
However, TF outliers and CTFR outliers do not always correspond.  In
particular:

\begin{itemize}

\item{Some kinematically anomalous galaxies are CTFR outliers, but not TF
outliers.  For example, the two CTFR outliers labeled N and O have very red
colors and truncated rotation curves.  Because these factors cause partially canceling
TF offsets, these galaxies fall within the general cloud of TF scatter in
spite of their kinematic abnormality.  However, their TF residuals are
actually incorrect for their colors, so they do not follow the CTFR
relation.  Likewise, galaxy P deviates slightly from the CTFR relation but
remains within the cloud of TF scatter.}

\item{The TF outliers labeled F and K in Fig.~\ref{fg:pairsalone} are
not CTFR outliers, nor are they flagged as kinematically anomalous.  Their
large TF residuals are actually in line with expectations based on their
extremely blue colors, so they appear to define a young-starburst extension
of the CTFR relation.  Whether their TF residuals are actually reliable is
unclear: indeed, galaxy K has a moderately high rotation curve asymmetry
that would have reached 10\% if we had adopted a higher-resolution
asymmetry measure (\S~\ref{sc:quant}).  These galaxies may have recently
relaxed onto the CTFR relation from an earlier state of more severe
kinematic disturbance.}

\end{itemize}

In addition to our rejection strategy, our method of measuring velocity
widths is also essential for defining a tight Close Pairs Survey CTFR
relation.  We use a robust velocity width measure, $V_{pmm}$, that yields
reliable TF offsets even when rotation curves are somewhat asymmetric,
below our 10\% rejection threshold.  K02 find that for disturbed or
otherwise non-canonical rotation curves, $V_{pmm}$ produces more reliable
results than $V_{2.2}$, the measure adopted by B01 following
\citet{courteau:optical}.  As a result, we find a more meaningful pattern
of TF outliers than B01.  Consider the eight TF outliers identified by B01
(labeled A--H in Fig.~\ref{fg:pairsalone}): (i) five are still TF outliers
in our analysis, and they are both kinematically anomalous and CTFR
outliers (C, D, E, G, and H); (ii) two are no longer strong TF outliers in
our analysis, and they are neither kinematically anomalous nor CTFR
outliers (A and B; notably these two are the only galaxies for which B01
could not find a physical basis for outlier behavior); (iii) one is still a
TF outlier, but its offset is consistent with a strong starburst that
follows the CTFR relation, and it is not flagged as kinematically
anomalous (F).  Using $V_{pmm}$ also reveals several new examples that
confirm the close correspondence between kinematic anomalies and CTFR
outliers, independent of whether a galaxy lies on the TF relation (I, J, K,
N, O, P).\footnote{\label{fn:pairsoutliers} Galaxies L and M deviate
slightly from the CTFR relation despite acceptable asymmetry and
truncation measures.  We suspect misleading photometric inclinations:
galaxy M is so distorted by interaction that its inclination is not well
defined, while galaxy L has an enhanced spiral arm that may be lifted out
of the plane of a more edge-on disk (morphological classification notes
courtesy R. A. Jansen).  Galaxy M also shows an unusual feature in its
rotation curve, with the appearance of a separate kinematic system on one
side (Fig~\ref{fg:rcegs}).}

\subsection{Comparison of the Close Pairs Survey and the NFGS}
\label{sc:combresults}

We now turn to a comparative analysis of the Close Pairs Survey and
the NFGS, to see whether the TF and CTFR relations for the Close Pairs
Survey reveal evidence of interaction-induced luminosity enhancements
relative to the general galaxy population \citep[represented by the
NFGS,][]{jansen.franx.ea:surface}.  Fig.~\ref{fg:pairsandnfgs} compares
the inverse-fit TF relations for the NFGS and the Close Pairs Survey.  The
formal fit results yield a slope difference, with a slope of 7.93$\pm$0.29
for the Close Pairs Survey (solid black line) and a steeper slope of
9.61$\pm$0.41 for the NFGS (solid gray line), excluding kinematically
anomalous galaxies in both cases.  The slope difference is formally
significant at 3.3$\sigma$ confidence.  The size and significance of this
difference agree with the results of B01 and may in part reflect enhanced
star formation at the faint end of the TF relation.

However, in the region of Fig.~\ref{fg:pairsandnfgs} where the NFGS
and the Close Pairs Survey overlap, the data look very similar (excluding
kinematically anomalous galaxies).  In fact, we find that the entire slope
difference comes from two sources: (1) galaxies F and K, and (2) the
difference between the two surveys' luminosity distributions
(Fig.~\ref{fg:twosampprops} and \S~\ref{sc:data}).  We argue below that
galaxies F and K may be the only galaxies in the Close Pairs Survey with
significant interaction-driven luminosity boosts that are not accompanied
by severe kinematic anomalies.  Without galaxies F and K, the Close Pairs
Survey TF relation would have a slope of 8.8 (dotted line).  Likewise, if
we weight each data point in the NFGS TF relation by the ratio of the two
surveys' luminosity distributions (i.e., the ratio of the histograms shown
in Fig.~\ref{fg:twosampprops}, but with kinematically anomalous galaxies
removed), the resulting TF fit yields a slope of 8.8.  We conclude that the
two surveys' TF relations are quite similar, except for a population of
disturbed galaxies within the Close Pairs Survey, which includes galaxies F
and K as well as the galaxies we have rejected because of severe kinematic
anomalies (not all of which are TF outliers).  We stress that this
disturbed population almost certainly reflects enhanced star formation from
interactions, consistent with B01.  However, quantitatively separating
luminosity and velocity offsets is impossible for most of these galaxies.

Nonetheless, we do see possible evidence for pure luminosity boosts in
the Close Pairs Survey, based on the position of galaxies F and K in
the CTFR relation.  To construct the CTFR relations for the Close
Pairs Survey and the NFGS in Fig.~\ref{fg:pairsandnfgs}, we compute
the TF residuals for both surveys relative to the NFGS TF relation,
which represents the TF relation for the general galaxy population.
Comparing the two CTFR relations, we find very similar locii, except
for an extension of the Close Pairs Survey CTFR relation toward bluer
colors.  Blueward of $(B-R)_e=0.6$, there are four Close Pairs Survey
galaxies, but no NFGS galaxies.  If the frequency of very blue
galaxies in the Close Pairs Survey were representative of the general
galaxy population, the probability of the NFGS containing zero very
blue galaxies would be 3\% in a random sample, implying that the
difference between the surveys is probably real.  Two of the very blue
Close Pairs Survey galaxies are kinematically anomalous, but the other
two (galaxies F and K) lie on the CTFR relation, extending it as far
as $(B-R)_e\sim0.3$.

These two galaxies' $\sim$2.5 mag TF residuals probably reflect luminosity
enhancements from starbursts.  We cannot rule out kinematic effects,
especially for galaxy K (\S~\ref{sc:pairsresults}), but the fact that both
galaxies have objectively acceptable kinematic anomalies and also fall on
the CTFR relation is reassuring. As argued by K02, the slope of the CTFR
relation in optical passbands is consistent with the slope expected when
the dominant physics determining TF offsets involves star formation.
Population synthesis models combined with a variety of galaxy formation
models all yield similar predictions for the slope of the relation between
color and stellar mass-to-light ratio \citep{bell.:stellar}, which may be
converted to predictions for the slope of the CTFR relation given certain
simplifying assumptions (e.g., that dark matter fractions are constant).
The predicted and observed slopes agree well, suggesting that star
formation explains most of the CTFR slope (K02).  In this context, we
interpret the blue extension of the Close Pairs CTFR relation as evidence
of very bright, young starbursts driven by interactions.  For reasonable
star formation histories, even 10\% mass starbursts generally extend rather
than depart from the CTFR relation \citep[][ see their
Fig.~5]{bell.:stellar}. After 1 Gyr, such bursts tend to fall slightly
above the CTFR relation, but still within the observational scatter set by
our measurement errors.

Younger bursts may show more extreme luminosity offsets that do not
follow the CTFR relation.  Unfortunately, most CTFR outliers have
kinematic anomalies, so we cannot readily disentangle luminosity and
velocity offsets for these galaxies.  Even for those few CTFR outliers
without kinematic anomalies, velocity offsets may play a role.  All
but one of the non-anomalous NFGS CTFR outliers labeled $w$--$z$ in
Fig.~\ref{fg:pairsandnfgs} have companions that could be causing
luminosity offsets, but these same companions may also cause velocity
offsets, for example via systematic inclination errors from
photometric distortion. In one case we also suspect a large asymmetry
hidden by a poorly sampled rotation curve.

K02 find that integrated H$\alpha$ equivalent widths \citep[i.e., with
the spectrograph slit scanned over the entire
galaxy][]{jansen.fabricant.ea:spectrophotometry} also correlate with
TF residuals for the NFGS, and Fig.~\ref{fg:bothew} shows that
galaxies $w$, $y$, and $z$ are outliers from the integrated
EW(H$\alpha$)--TF residual relation just as they are from the
color--TF residual relation.  However, the integrated
EW(H$\alpha$)--TF residual relation is more scattered than the CTFR
relation, and the status of its outliers is less obvious. Moreover,
the use of central rather than integrated EW(H$\alpha$) measurements
degrades the relation considerably (compare both panels of
Fig.~\ref{fg:bothew}).  Unfortunately, central measurements are the
only type available for the Close Pairs Survey.  Although some CTFR
outliers remain outliers in the central EW(H$\alpha$)--TF residual
relation, the latter relation does not offer a clean way to separate
offsets affected by kinematic anomalies from reliable luminosity
boosts.

Most blue starburst galaxies in the Close Pairs Survey are fainter than
M$_{\rm B}\sim-21$, despite the overall survey bias toward bright galaxies
(e.g., Fig.~\ref{fg:pairsalone}).  The NFGS shows a similar tendency,
resulting in asymmetric scatter at the bright end of the TF relation
(K02).  The absence of very blue galaxies at high luminosities may reflect
a hierarchical formation history in which bright galaxies form early via
mergers that consume most of the available gas, so that in later
interactions these galaxies lack the fuel necessary for major starbursts.
B01 show that the Close Pairs data are consistent with interaction-driven
starbursts of relatively constant size, so that the fractional contribution
of the starburst light to the total light is negligible for the largest
galaxies.

\section{The Origins of Kinematic Anomalies}
\label{sc:interp}

Kinematically anomalous galaxies with severely truncated and/or asymmetric
rotation curves represent nearly 20\% of the Close Pairs Survey TF sample,
roughly twice the frequency seen for the corresponding NFGS sample.  The
bulk of the difference arises from the 3--4$\times$ higher incidence of
strong rotation curve asymmetries in the Close Pairs Survey (10/88 vs.\
2/73).  Truncated rotation curves also occur more often in the Close Pairs
Survey (9/88 vs.\ 5/73), but with only marginal statistical significance.
The significance of the frequency difference for strong rotation curve
asymmetries depends upon interpretation.  A K-S test finds no significant
difference between the two continuous asymmetry distributions shown in
Fig.~\ref{fg:asymdist}.  However, accepting that asymmetries above 8--10\%
are ``anomalous,'' i.e., represent a discontinuously disturbed population,
the difference in the rate of anomalies between the two surveys is highly
significant.  For example, if we assume that the NFGS reflects the
underlying parent distribution of asymmetries, the probability of obtaining
rotation curve asymmetries $>$10\% in 10/88 Close Pairs Survey galaxies is
$3.6\times10^{-7}$.  In reality, systematics dominate this problem, and
statistical tests are of limited utility.  Confirmation with other samples
would be more valuable.

Most of our kinematically anomalous galaxies have either strongly
asymmetric or truncated rotation curves, but not both.  However,
rotation curve truncation may prevent detection of non-central
asymmetries (\S~\ref{sc:asym}).  Fig.~\ref{fg:pairsandnfgs} shows that
large rotation curve asymmetries generally occur in blue galaxies,
while truncated rotation curves generally occur in red galaxies
(though the latter may have relatively blue centers,
\S~\ref{sc:srctrunc}).

\subsection{Sources of Rotation Curve Asymmetries}

The fact that strong rotation curve asymmetries occur more often in the Close Pairs
Survey than in the NFGS provides circumstantial evidence that close
neighbors play a role in driving such asymmetries.  In their selection
criteria, the two surveys differ primarily in local environmental distribution
(pairs vs.\ any environment) and luminosity distribution (top-heavy vs.\
representing a broad range of luminosities, Fig.~\ref{fg:twosampprops}).
If anything, the difference in luminosity distributions causes us to
underestimate the difference in rotation curve asymmetries: the top-heavy luminosity
distribution of the Close Pairs Survey should not favor large rotation
curve asymmetries, because luminosity and rotation curve asymmetry anticorrelate in the
general galaxy population (K02 and NFGS symbols in
Fig.~\ref{fg:asymdist}b).  We conclude that the higher rate of strong rotation curve
asymmetries in the Close Pairs Survey very likely reflects the selection of
close pair environments, barring any systematic difference in the data
(unlikely given our conclusions regarding resolution dependence and
reproducibility in \S~\ref{sc:asym}--\ref{sc:repro}).

Interactions with neighbors or generally dense environments may induce
rotation curve asymmetries via several mechanisms.  All rotation curves
show some degree of asymmetry, which correlates with inclination, Hubble
type, and galaxy color\footnote{The rotation curve asymmetry--TF residual
correlation reported by K02 is related to the color--rotation curve
asymmetry correlation in Fig.~\ref{fg:colorasym} via the much stronger
color--TF residual correlation.}  (Fig.~\ref{fg:colorasym};
\citeauthor{kannappan.fabricant:kinematic} 2001;
\citeauthor{kannappan:kinematic} 2001; see also
\citeauthor{beauvais.bothun:precision} 1999) and may be related to small
satellite interactions or turbulence from self-regulated star formation.
Strong asymmetries may arise from tidally disrupted gas and/or decoupling
between the gas dynamical center of the galaxy and the stellar continuum
peak position during violent encounters
\citep[e.g.,][]{mihos:development,barton.geller.ea:tully-fisher,kornreich.lovelace.ea:sloshing}.
Asymmetries may also reflect the presence of multiple gas components due to
merging systems or strong bars, since standard rotation curve analysis
techniques assign a single velocity per spatial position even when there
are multiple velocity components, and which velocity component dominates
the result may vary strongly with local changes in emission-line strength
\citep{barton.kannappan.ea:rotation}.  \citet{rubin.waterman.ea:kinematic}
find evidence that cluster interactions may produce long-lived disturbances
in rotation curves \citep[but see][]{dale.giovanelli.ea:signatures}; we
note that our surveys contain few galaxies in dense cluster environments,
because field environments are statistically more common
(Fig.~\ref{fg:twosampprops}).  Extinction in dusty starbursts can also
cause rotation curve asymmetries, although in our surveys, most galaxies
with large rotation curve asymmetries show blue colors atypical of dusty
systems (Fig.~\ref{fg:pairsandnfgs}).

We plan to investigate how these mechanisms combine to explain observed
rotation curve asymmetries in a future paper using integral-field
kinematics (with M. Bershady, in preparation).  In the absence of
understanding the origin of rotation curve asymmetries, we cannot at
present explain their relationship to TF offsets.  The large TF offsets we
observe for galaxies with high rotation curve asymmetries may reflect
concomitant luminosity enhancements and/or symmetric rotation curve
distortions \citep[]{barton.bromley.ea:kinematic}, in addition to the
direct effects of rotation curve asymmetries.

\subsection{Sources of Rotation Curve Truncation}
\label{sc:srctrunc}

The slight difference in rotation curve truncation frequency between the NFGS and
the Close Pairs Survey may reflect nothing more profound than S/N.  If
we edit each NFGS rotation curve to remove individual points that
would be lost at the S/N of the Close Pairs Survey (using the relative
S/N determined by comparing galaxies common to the two surveys in
\S~\ref{sc:repro}), three additional NFGS rotation curves become truncated, so
that the frequency of truncation is indistinguishable between the
surveys.\footnote{One might also worry about the larger mean redshift
of the Close Pairs Survey; however, rotation curves depend on surface
brightness which is distance-invariant at low $z$, and in any case the
median apparent magnitude for the Close Pairs Survey galaxies with truncated
rotation curves is only 0.5 mag fainter than the median apparent magnitude for the
NFGS.  We see no evidence for systematically poor weather conditions
during observations of galaxies with truncated rotation curves.}

However, physical arguments predict an enhanced rotation curve truncation
rate for galaxies selected in pairs (as discussed by B01).  Neighbor
interactions can trigger disk gas inflow and central starburst activity
\citep[e.g.,][]{hernquist.mihos:excitation,barton-gillespie.geller.ea:tidally},
causing optical emission lines to be preferentially detected at small
radii.  In addition, galaxies in cluster environments may show radially
truncated emission due to gas stripping or tidal harassment
\citep{gunn.gott:on,moore.lake.ea:morphological}, though only a few
galaxies in the Close Pairs Survey have sufficiently dense environments for
cluster mechanisms to play a role (and in particular, the galaxies with
truncated rotation curves have modest densities on the scale of
Fig.~\ref{fg:twosampprops}f, ranging from $-$0.2 to 0.5 on the log scale,
with median $-$0.06).

These physical processes might have produced more truncated rotation curves
in the Close Pairs Survey if the survey were not deficient in
low-luminosity galaxies.  Even with this deficiency, truncated rotation
curves appear more common at lower luminosities within the survey
(Fig.~\ref{fg:pairsalone}; the luminosity distributions for truncated and
non-truncated rotation curves differ at 95\% confidence in a K-S test).
Moreover, {\em physically induced} rotation curve truncation may also be
more common at lower luminosities, based on a two-step chain of inference.
(1) Several Close Pairs Survey galaxies with truncated rotation curves have
unusual color profiles suggestive of gas inflow processes: their colors
within $r_e$ are bluer than their outer disk colors.  In general,
blue-centered galaxies include blue compact morphologies, such as
blue-centered emission-line S0 galaxies, as well as many later-type
morphologies, and they often show evidence of interactions and mergers
\citep{kannappan.jansen.ea:forming}.  Statistically, truncated rotation
curves in the Close Pairs Survey correlate with blue-centered galaxies at
98\% confidence (four of nine galaxies).  (2) Blue-centered galaxies are
nearly always fainter than M$_{\rm B}=-20$
\citep{kannappan.jansen.ea:forming}.  Thus rotation curve truncation
associated with blue-centered galaxies will occur primarily at low
luminosities.  Because low-luminosity galaxies are underrepresented in the
Close Pairs Survey compared to the NFGS (\S~\ref{sc:data}), this trend
probably weakens any statistical difference between the two surveys' rates
of rotation curve truncation, insofar as that difference is related to
interaction-driven gas inflow.  Therefore in our view interactions remain a
likely source of rotation curve truncation.

\section{Implications for High-$z$ TF Studies}
\label{sc:hizsec}

The pattern of TF offsets for kinematically anomalous galaxies in the Close
Pairs Survey looks very similar to the ``differential luminosity
evolution'' seen in some high-$z$ TF studies
\citep[e.g.,][]{simard.pritchet:internal,ziegler.b-ohm.ea:evolution}.
Fig.~\ref{fg:zieglercmp} shows the Close Pairs Survey and the NFGS
alongside high-$z$ TF data from the FORS Deep Field \citep[the survey used
by][]{ziegler.b-ohm.ea:evolution}, courtesy of A. B\"ohm
\citep{b-ohm.ziegler.ea:tully-fisher}.
\citet{b-ohm.ziegler.ea:tully-fisher} derive velocities by fitting
simulated RCs, including slit width and seeing effects, to observed
(extracted) one-dimensional RCs.  For both the high-$z$ sample and the
Close Pairs Survey, the largest TF offsets occur in the same region of
parameter space, i.e., at luminosities fainter than M$_{\rm B}\sim-21$ and
velocity widths less than $\log{2V}\sim2.2$.  In both cases, this pattern
of TF offsets creates a shallow TF slope compared to the reference slope
defined by the NFGS.

Furthermore, at both high and low $z$, this pattern of offsets is linked to
kinematic anomalies.  In the Close Pairs Survey, the largest TF offsets
correspond to kinematically anomalous galaxies, which have offsets of up to
4 mag.  In the FORS Deep Field sample, much of the slope evolution is
driven by ``low-quality'' data points, identified as such by B\"ohm et al.\
because the corresponding rotation curves ``have a smaller radial extent
and partly feature signatures of moderate kinematic perturbations like
waves or asymmetries'' \citep[][ data table notes; no quantitative criteria
given]{b-ohm.ziegler.ea:tully-fisher}.  By analogy with the Close Pairs
Survey, we suggest that the slope evolution in the FORS Deep Field may
reflect kinematic anomalies caused by companions or minor mergers. The
frequency of interactions and mergers is expected to increase with $z$
\citep{patton.pritchet.ea:dynamically,murali.katz.ea:growth}, and
physically significant interactions are not always visually obvious. In
fact, many kinematically anomalous galaxies in the NFGS show only subtle
interaction evidence, with faint or already merging small companions.
Conversely, large, obvious galaxy pairs need not be influenced by
interactions.\footnote{These two considerations probably largely explain
why B\"ohm et al.\ do not find a strong link between known pairs and large
TF offsets.  We note however that in their low-quality subsample, which is
identified based on kinematic anomalies and drives most of the slope
evolution, they do report a slight overrepresentation of pair/cluster
candidates among galaxies with large TF offsets.} The brightest galaxies in
the Close Pairs Survey all have companions (by definition), yet they
display minimal starburst activity and few strong kinematic anomalies,
possibly because the galaxy formation process consumes most of the gas in
massive galaxies early, inhibiting gas-dynamical processes in later
interactions (\S~\ref{sc:combresults}).

Interaction-induced kinematic anomalies may be more common at high $z$ than
is generally recognized.  Rotation curve asymmetries like those in
Fig.~\ref{fg:rcegs} will not always be obvious at the resolution of
high-$z$ data.  Moreover, the high frequency of blue compact galaxies in
many high-$z$ TF studies probably reflects interaction-driven starburst
activity that can cause rotation curve truncation via gas inflow processes
\citep[B01 and][]{barton.:possible}.  Low S/N is another potential source
of rotation curve truncation and thereby TF outliers in some high-$z$
studies, especially studies already biased toward centrally concentrated
star formation \citep[as][~suggests for the Simard \& Pritchet 1998 study;
see also Kobulnicky \& Gebhardt 2000]{kannappan:kinematic}.

To the extent that kinematically anomalous galaxies drive the apparent
luminosity evolution in high-$z$ TF studies, the nature of that evolution
is unclear.  At present, disentangling luminosity and velocity offsets for
galaxies with rotation curve asymmetries is not possible.  Estimating
velocity offsets for galaxies with rotation curve truncation is also
difficult, although we do know that for blue compact galaxies in
particular, underestimated velocity widths may yield average TF offsets of
order 2 mag for unresolved optical data \citep[based on studies of their
low-$z$ analogues,][]{barton.:possible,pisano.kobulnicky.ea:gas}.

The exact effects of rotation curve truncation may differ at low and high
$z$, because most high-$z$ studies do not compute velocity widths directly
from rotation curves but instead analyze kinematic and photometric profiles
together
\citep[e.g.,][]{vogt.forbes.ea:optical,simard.pritchet:analysis,ziegler.b-ohm.ea:evolution}.
These modeling techniques generally assume that emission-line flux and
disk-continuum flux profiles are simply related by scalings in radial
extent and intensity.  The models further assume a basic form for the
rotation curve constrained by exponential fits to the spatial flux
distribution, even when HST data are available to model bulge components in
the images.  Within these assumptions, using high-$z$ techniques may
mitigate truncation-induced velocity offsets in the TF relation (although
probably not asymmetry-induced velocity offsets).  However, the presence of
bulges, bars, central starbursts, and morphological distortions will
severely compromise such modeling.  A realistic evaluation of the effects
of rotation curve truncation on high-$z$ data will require simulating
high-$z$ resolution, S/N, and analysis techniques using a low-$z$ sample
like the NFGS that includes all Hubble types, as well as barred, peculiar,
and interacting galaxies.

In short, the contribution of velocity offsets to high-$z$
luminosity evolution is uncertain but potentially large.
Discrepancies between high-$z$ TF studies that report minimal
evolution
\citep[e.g.,][]{vogt.phillips.ea:optical,bershady.haynes.ea:rotation},
and 1--2 mag evolution
\citep[e.g.,][]{rix.guhathakurta.ea:internal,simard.pritchet:internal,ziegler.b-ohm.ea:evolution}
may in part reflect differences in sample selection criteria that lead
to higher or lower percentages of kinematically anomalous galaxies.
For example, the \citet{vogt.phillips.ea:optical} sample favors large,
undisturbed disks, while the \citet{ziegler.b-ohm.ea:evolution} sample
includes all types of elongated emission-line galaxies, and thus
almost certainly galaxies with kinematic anomalies.

Measuring luminosity evolution reliably will require identifying and
rejecting kinematically anomalous galaxies in high-$z$ samples and/or
accounting for the velocity offsets inherent in these galaxies' TF offsets.
Isolating TF evolution associated with kinematic anomalies may also be
interesting in its own right, as a way to probe the evolving role of
mergers and interactions as a function of luminosity and redshift.  At low
$z$, the NFGS shows a distinct population of kinematically anomalous
galaxies fainter than M$_{\rm B}=-18$ (Fig.~\ref{fg:zieglercmp}), perhaps
reflecting late-epoch galaxy formation activity on the smallest mass
scales.  Below we consider four possible strategies for analyzing high-$z$
TF data in the presence of kinematic anomalies.

\subsection{Identification Based on Rotation Curve Properties}

By definition, the most accurate way to identify kinematically
anomalous galaxies is via their rotation curve properties.  Spatially
resolved rotation curves combined with good $r_e$ measurements should
be sufficient to flag cases of rotation curve truncation, using a
rejection threshold optimized for the TF sample under study
(\S~\ref{sc:trunc}).  However, high-$z$ kinematic data generally lack
the spatial resolution necessary to measure reliable rotation curve
asymmetries (\S~\ref{sc:asym}).  A new generation of 20--30-meter
telescopes with adaptive optics may enable such measurements in the
future.

\subsection{Identification Based on Morphology}

Most low-$z$ TF studies would reject the strongest TF outliers in the
Close Pairs Survey on morphological grounds.  Of the nine strongest TF
outliers in the survey (TF residuals brighter than $-$1.5 mag in
Fig.~\ref{fg:pairsalone}), three are emission-line S0 galaxies,
while the other six are distorted by interactions, including the two that
follow the CTFR relation.  High-$z$ TF studies typically compare against
low-$z$ calibration samples that exclude such early-type or disturbed
morphologies.  However, the high-$z$ samples themselves may not exclude
such morphologies, due to sample selection procedures that rely on
spectroscopic galaxy types and/or low physical-resolution images.  For
example, the abundance of compact narrow emission line galaxies
\citep[CNELGs,][]{guzman.koo.ea:on} in high-$z$ samples may in part reflect
the inclusion of emission-line S0 galaxies. \citet{barton.:possible} show
that four TF outliers in the Close Pairs Survey have properties that
suggest they are counterparts to CNELGs at higher redshift; of these four,
two are emission-line S0's and one is an early-type peculiar galaxy
\citep[according to the classifications
of][]{kannappan.jansen.ea:forming}. Emission-line S0 galaxies form a
prominent subpopulation of kinematically anomalous galaxies in the NFGS as
well, as indicated by the asterisks in Fig.~\ref{fg:zieglercmp}.

In general, all high-$z$ TF studies that report differential
luminosity evolution employ selection criteria that allow
emission-line S0 and disturbed spiral morphologies.  With HST
imaging, a strategy of rejecting both classes of galaxy might
successfully eliminate many kinematic anomalies at high $z$ and
establish whether the non-anomalous galaxy population shows
differential luminosity evolution.  However, some anomalous
galaxies would probably escape rejection, while some non-anomalous
galaxies would probably be thrown out, possibly including galaxies
essential to measuring luminosity evolution like the two galaxies that
seem to extend the CTFR relation for the Close Pairs Survey.  Also,
this approach would be too imprecise to support a detailed analysis of
kinematically anomalous galaxies for their own sake.

\subsection{Identification Based on the Color--TF Residual Relation}

If a color--TF residual relation can be established at high $z$,
then identifying kinematically anomalous galaxies based on their close
correspondence with CTFR outliers may be easier than identifying them
from their rotation curve properties, with comparable effectiveness
for separating reliable luminosity evolution from the ambiguous TF
offsets of kinematically anomalous galaxies.  In the Close Pairs
Survey, reliable luminosity offsets lie along the CTFR relation,
possibly extending it toward bluer colors, while offsets that do not
follow the CTFR relation nearly always correspond to kinematic
anomalies.

With a large enough sample, it may be possible to establish a
high-$z$ CTFR relation just using sigma clipping and rejection based
on morphology or rotation curve truncation.
\citet{bershady.haynes.ea:rotation} report initial evidence for a
high-$z$ CTFR relation based on a sample of disk galaxies
spanning a broad range of colors.  The low-$z$ CTFR relation may be
used to help refine the locus of the high-$z$ CTFR relation, assuming
closely standardized analysis techniques, though we caution that the
CTFR relation may evolve. If defining a tight high-$z$ relation proves
difficult, establishing its locus might require a small sample of
galaxies with well-resolved rotation curves, free of kinematic
anomalies according to the criteria presented here.  Obtaining such
data at high $z$ would probably require deep ground-based spectroscopy
with adaptive optics.  Fortunately, once a high-$z$ CTFR
relation has been established, kinematic anomalies can be eliminated
from a larger sample with unresolved or poorly resolved rotation
curves simply by rejecting CTFR outliers.  Rejecting CTFR outliers can
also help to eliminate galaxies with faulty inclination estimates or
otherwise spurious data.  Any reliable luminosity evolution that
remains can then be measured by comparing the high- and low-$z$ CTFR
relations, perhaps with the precaution of obtaining well-resolved
rotation curves for any high-$z$ galaxies that extend the CTFR
relation.

Since measuring equivalent widths is often easier than measuring
colors at high $z$, attempting to identify kinematic anomalies using
the EW(H$\alpha$)-- or EW([OII])--TF residual relations might also be
worthwhile.  These relations have already proven useful for estimating
bulk luminosity shifts related to differences in mean emission-line
strength between high- and low-$z$ TF samples
\citep{kannappan:kinematic,kannappan.gillespie.ea:interpreting,kannappan.fabricant.ea:calibrating}.
However, the emission-line strength--TF residual relations are noisier than the CTFR
relation even at low $z$ (K02), and equivalent widths can depend
strongly on spectroscopic aperture
\citep[\S~\ref{sc:combresults}, ][]{jansen.fabricant.ea:spectrophotometry}.
Thus it is not yet clear that emission-line strength can serve as a
surrogate for color for identifying kinematically anomalous galaxies.

\subsection{Construction of Matching Low-$z$ Calibration Samples}

An alternative to rejection at high redshift is greater inclusion at low
redshift.  For example, the NFGS includes kinematically anomalous galaxies
similar to those at high $z$, because it is a statistically representative,
morphology-blind sample of the local galaxy population.  Many of the worst
TF outliers in the NFGS are emission-line S0 or irregular late-type
galaxies with kinematic oddities such as large rotation curve asymmetries,
truncated rotation curves, or counterrotating gas and stars (K02; see also
\citeauthor{kannappan.fabricant:broad} 2001).  Including such kinematically
anomalous galaxies in determining the low-$z$ reference TF relation could
help to eliminate apparent evolution between low and high redshift that
really reflects different sample selection criteria.

To first order, an all-inclusive low-$z$ reference sample should provide a
good calibration for a magnitude-limited high-$z$ sample like the FORS Deep
Field \citep{b-ohm.ziegler.ea:tully-fisher}.  To illustrate, we compare TF
fits for the NFGS, the FORS Deep Field, and the Close Pairs Survey in
Fig.~\ref{fg:zieglercmp}.  The NFGS defines the reference relation, shown
as a solid line in all three panels.  The dashed lines show the best-fit TF
relations for the other two samples with the slope held fixed to the NFGS
value, where we include two lines for the FORS Deep Field sample to show
fits with and without the data designated as ``low-quality'' by
\citet{b-ohm.ziegler.ea:tully-fisher} because of kinematic perturbations or
inadequate rotation curve extent.  While we cannot directly assess whether
B\"ohm et al.'s low-quality data points would meet our quantitative
criteria for kinematic anomalies, the 0.18 mag difference between fits with
and without these low-quality points gives some idea of the likely
contribution of kinematically anomalous galaxies to the TF offset for the
FORS Deep Field.  This contribution is quite similar to the 0.16 mag offset
for the Close Pairs Survey relative to the NFGS.  Apart from the 0.18 mag
attributed to low-quality points, we find an offset of only 0.28 mag for
the FORS Deep Field relative to the NFGS.
\citet{b-ohm.ziegler.ea:tully-fisher} find a larger mean offset in part
because they follow other high-$z$ workers in using the TF relation of
\citet{pierce.tully:luminosity-line} as a low-$z$ reference relation.  This
relation has a known zero-point error of $\sim$0.3--0.4 mag \citep[][ page
776]{tully.pierce:distances}, depending on whether one prefers H$_0$ = 75
or 70 \kmsMpc.  A high-$z$ TF offset of $\sim$0.4 mag would be expected for
the FORS Deep Field based on the sample's bluer mean color compared to the
NFGS, assuming the same CTFR relation observed at low $z$ (A. B\"ohm,
private communication).  Thus the observed $\sim$0.3 mag offset is
consistent with a combination of evolution in mass-to-light ratio (as
suggested by B\"ohm et al.) and perhaps some evolution in stellar-to-total
mass fraction \citep[as discussed for the Vogt et al.\ 1997 sample
in][]{kannappan.gillespie.ea:interpreting}.  However, uncertainties and
systematics dominate at this level: (i) scatter in the high-$z$ data, (ii)
differences between velocity measures \citep[e.g., using a $V_{pmm}$
equivalent at high $z$ could decrease the evolution by $\sim$0.1 mag from
what we measure, ][]{kannappan:kinematic}, (iii) differences in Galactic
extinction corrections\footnote{We standardize internal but not Galactic
extinction corrections \citep[the FORS Deep Field corrections are described
in][]{heidt.appenzeller.ea:fors}. Applying our Galactic extinction
conventions to the high-$z$ data would increase the evolution by
$\sim$0.04--0.08 mag from what we measure, depending on the passband
closest to rest-frame B for a given galaxy \citep[based on known zero point
differences between different extinction
maps,][]{burstein:line-of-sight}.}, (iv) the unknown true frequency of
kinematic anomalies at high $z$, and (v) differences in color or
surface-brightness selection biases, which may affect the relative
frequency of blue colors and/or kinematic anomalies (e.g., by favoring high
surface brightness blue compact galaxies).

In addition to affecting mean offsets, kinematic anomalies also affect
slope evolution.  As previously discussed, the FORS Deep Field TF relation
is shallower than the NFGS TF relation (inverse-fit slopes of $-6.3$ and
$-7.4$, respectively; the high-$z$ fit is shown as a dotted line
in Fig.~\ref{fg:zieglercmp}). This result confirms the slope evolution
reported by \citet{ziegler.b-ohm.ea:evolution}, although the exact slope
differs because of different extinction corrections and fitting methods.
Unlike B\"ohm et al., however, we do not find statistically significant
slope evolution for the high-quality points taken alone. Separate inverse
fits to the high and low quality data in Fig.~\ref{fg:zieglercmp} suggest
that the shallower slope is produced almost entirely by the low-quality
subsample, which is subject to kinematic anomalies. Specifically, the slope
evolution is driven by anomalous galaxies with large TF offsets of up to
$\sim$2 mag at intermediate luminosities, M$_{\rm B}\sim-18$ to $-21$
(Fig.~\ref{fg:zieglercmp}).  By comparison, NFGS galaxies with strong
kinematic anomalies and large TF offsets typically have luminosities
fainter than M$_{\rm B}=-18$.
\citet[][]{mall-en-ornelas.lilly.ea:internal} report an analogous
evolutionary shift in the characteristic luminosities of blue compact
galaxies from high to low $z$.  These trends may reflect mass-dependent
evolution in the rate of starbursts and gas-dynamical disturbances driven
by galaxy mergers and interactions, following the hierarchical tendency for
today's more massive galaxies to show peak formation activity at higher
redshifts than today's smaller galaxies
\citep[e.g.,][]{cowie.songaila.ea:new}.  In this view, both the number
density and the luminosity distribution of kinematically anomalous galaxies
would be expected to evolve, and {\em the optimal low-$z$ calibration
sample for isolating luminosity offsets from velocity offsets would be a
sample that simulated the expected luminosity distribution of kinematic
anomalies at higher $z$.}  The simulation would ideally involve adjusting
both the interacting galaxy distribution and the data quality (e.g.,
rotation curve S/N, inclination errors) to match the high-$z$ sample under
study.  Modeling high-$z$ analysis techniques would also be essential, and
such modeling might lead to development of a high-$z$ analogue of $V_{pmm}$
that would minimize velocity offsets from minor kinematic distortions.

\section{Conclusions}

We have demonstrated robust methods for measuring luminosity evolution in
TF samples with a high frequency of rotation curve anomalies, such as might
be expected at high redshift.  The Close Pairs Survey of
\citet{barton.geller.ea:tully-fisher} is ideal for this analysis, as a
low-$z$ TF sample with high-quality data and many similarities to high-$z$
TF samples: optical emission-line rotation curves, morphology-blind
selection, and a large number of interacting galaxies.  The Nearby Field
Galaxy Survey
\citep[NFGS,][]{jansen.franx.ea:surface,kannappan.fabricant.ea:physical}
offers a low-$z$ reference sample with similar features, but with a more
typical number of interacting galaxies.  We have extended
\citeauthor{barton.geller.ea:tully-fisher}'s previous TF analysis of the
Close Pairs Survey, which showed that both starbursts and kinematic
disturbances can create apparent ``luminosity evolution'' for galaxies in
interacting pairs, by our demonstration of methods for isolating
potentially spurious luminosity offsets associated with severe kinematic
anomalies from reliable luminosity offsets clearly linked to star
formation.

The largest apparent luminosity offsets in the Close Pairs Survey TF
relation correspond to galaxies with severe kinematic anomalies
\citep[asymmetric rotation curve shapes and/or radially truncated rotation
curve extents, using objective measures adapted
from][]{kannappan.fabricant.ea:physical}.  The pattern of these galaxies'
TF offsets looks much like the differential luminosity evolution claimed in
many high-$z$ studies, with the largest TF offsets at luminosities fainter
than M$_{\rm B}\sim-21$.  Excluding the galaxies with asymmetric or
truncated rotation curves, however, and adopting a robust velocity width
measure insensitive to minor kinematic distortions, we find that the TF
relations for the Close Pairs Survey and the NFGS are very similar, with no
significant evidence for overall luminosity enhancement in paired galaxies
relative to the general population.

Nonetheless, we do find evidence for luminosity enhancement when we compare
the color--TF residual (CTFR) relations for the two surveys.  Two galaxies
that are not objectively flagged as kinematically anomalous extend the CTFR
relation to very blue colors and large luminosity offsets, apparently
reflecting interaction-induced star formation.  Of course, kinematically
anomalous galaxies probably experience luminosity boosts as well, and in
fact two such galaxies are also exceptionally blue.  However, anomalous
galaxies are typically outliers from the CTFR relation, and their TF
offsets may include velocity offsets.  Unfortunately, it is presently
impossible to separate luminosity and velocity offsets for these galaxies.

If, as expected from hierarchical merging scenarios, the galaxy
interaction rate was higher in the past, kinematic anomalies may pose
a serious problem for high-$z$ TF studies.  Severe anomalies are
roughly twice as common in the Close Pairs Survey as in the NFGS
($\sim$20\% vs.\ $\sim$10\% of galaxies brighter than
$M_B=-18$). Galaxy interactions probably explain the 3--4 times higher
rate of strong rotation curve asymmetries in the Close Pairs Survey
compared to the NFGS.  The externally triggered gas inflow associated
with interactions can also lead to centrally concentrated line
emission and thereby rotation curve truncation, a problem that may be
compounded by low S/N data.  However, our data are inconclusive as to
the primary source of rotation curve truncation in the Close Pairs
Survey. To the extent that gas inflow processes play a role at high
$z$, some of the assumptions inherent in high-$z$ rotation-curve
fitting techniques may break down, leading to artificially low
velocity widths, as observed at low $z$.

The frequency of kinematic anomalies at high $z$ that would meet our
criteria is presently unknown.  We have shown that TF outliers associated
with kinematic anomalies in the Close Pairs Survey occupy the same part of
TF parameter space as the galaxies responsible for TF slope evolution in
some high-$z$ studies
\citep[e.g.,][]{simard.pritchet:internal,ziegler.b-ohm.ea:evolution}.
These studies tend to have morphology-blind selection criteria that would
include kinematically anomalous galaxies, which often have blue compact,
emission-line S0, or peculiar morphologies.  In contrast, studies that show
less slope evolution \citep[e.g.,][]{vogt.phillips.ea:optical} tend to
favor large disks and probably contain fewer kinematic anomalies.

We have also shown that the slope evolution in the high-$z$ FORS Deep
Field sample \citep{b-ohm.ziegler.ea:tully-fisher} is largely driven
by ``low-quality'' data points, labeled as such by B\"ohm et al.\
based on perturbations or limited radial extent in the rotation
curves.  Using the FORS Deep Field data, we find that most or all of
the TF evolution measured at high $z$ can be modeled as an overall
$\sim$0.3 mag luminosity offset at fixed slope, consistent with
evolution along the CTFR relation, plus a differential evolution
component associated with kinematically anomalous galaxies, which show
offsets as large as $\sim$2 mag at low luminosities but add only a
small $\sim$0.2 mag enhancement to the total TF offset for the
survey.  We note that the use of the outdated
\citet{pierce.tully:luminosity-line} TF calibration \citep[superseded
by][]{tully.pierce:distances} as a low-$z$ reference relation
contributes 0.3--0.4 mag of spurious luminosity evolution to many
high-$z$ TF studies.  At present, only the $\sim$0.3 mag offset
consistent with the CTFR relation can be reliably interpreted as
luminosity evolution.

TF slope evolution associated with kinematic anomalies may be interesting
for its own sake, as a source of data on mass-dependent evolution in the
frequency of mergers and interactions (or the frequency of gas-dynamical
disturbances caused by these events).  Consistent with mass-dependent
evolutionary trends in star formation histories and the luminosity function
\citep[e.g.,][]{cowie.songaila.ea:new}, kinematically anomalous galaxies in
the NFGS tend to be dwarf galaxies, fainter than M$_{\rm B}=-18$, while
analogous galaxies at high $z$ can be as bright as M$_{\rm B}\sim-21$.

We have considered four strategies for isolating reliable luminosity
offsets from offsets possibly associated with kinematic anomalies at
high $z$: identification of anomalies based on rotation curve
properties, identification based on morphology, identification based
on the color--TF residual relation, and inclusion of anomalies in
optimally matched low-$z$ calibration samples that reproduce the
distribution of anomalies expected at high $z$ (as well as high-$z$
selection criteria, data quality, and analysis techniques).  {\em The
color--TF residual relation may offer the simplest and most powerful
tool currently available for measuring luminosity evolution
independent of kinematic anomalies at high $z$,} especially when
combined with optimal low-$z$ calibration samples.  Unreliable TF
offsets associated with kinematic anomalies are typically CTFR
outliers.  Conversely, reliable luminosity enhancements lie on the
CTFR relation and extend it toward bluer colors.  Preliminary evidence
for a CTFR relation at high $z$ has already been reported
\citep{bershady.haynes.ea:rotation}.  If the high-$z$ CTFR relation
proves as tight as the Close Pairs Survey CTFR relation, then
identifying CTFR outliers will serve as the preferred method for
isolating kinematic anomalies in studies of luminosity evolution.

Once established, the high-$z$ CTFR relation may be applied to measuring
not only luminosity evolution but also the evolution of stellar populations
and stellar-to-total mass fractions
\citep[][]{kannappan.gillespie.ea:interpreting}.  The CTFR relation and the
analogous relations for EW(H$\alpha$) and EW([OII]) can also be used to
reconcile discrepancies between high-$z$ TF studies with different
selection biases in color or emission-line strength
\citep{kannappan:kinematic,kannappan.gillespie.ea:interpreting,kannappan.fabricant.ea:calibrating}.
Matching selection criteria at low and high $z$ is only the first step,
however, because of the potential for luminosity-dependent evolution in the
frequency of kinematic anomalies.  By combining well-matched low-$z$
calibration samples with careful modeling of kinematic anomalies and the
CTFR relation, future high-$z$ TF studies should be able to properly
account for the major uncertainties of existing studies and reach consensus
on how galaxy luminosities have evolved over cosmic time.

\acknowledgements Data for the FORS Deep Field were kindly provided
by A. B\"ohm, who also offered useful commentary on those data.  We
thank Niv Drory and David Koo for valuable feedback on this work, and
Karl Gebhardt, Margaret Geller, and an anonymous referee for helpful
comments on the manuscript.  We also thank Dan Fabricant and Margaret
Geller for the use of unpublished data from the NFGS and the Close
Pairs Survey.  EJB acknowledges support from NASA through Hubble
Fellowship grant HST-HF-01135.01, awarded by STScI, which is operated
by AURA for NASA under contract NAS 5-26555.  This research has used
NASA's Astrophysics Data System Bibliographic Services.

\newpage
\epsscale{1.}
\plotone{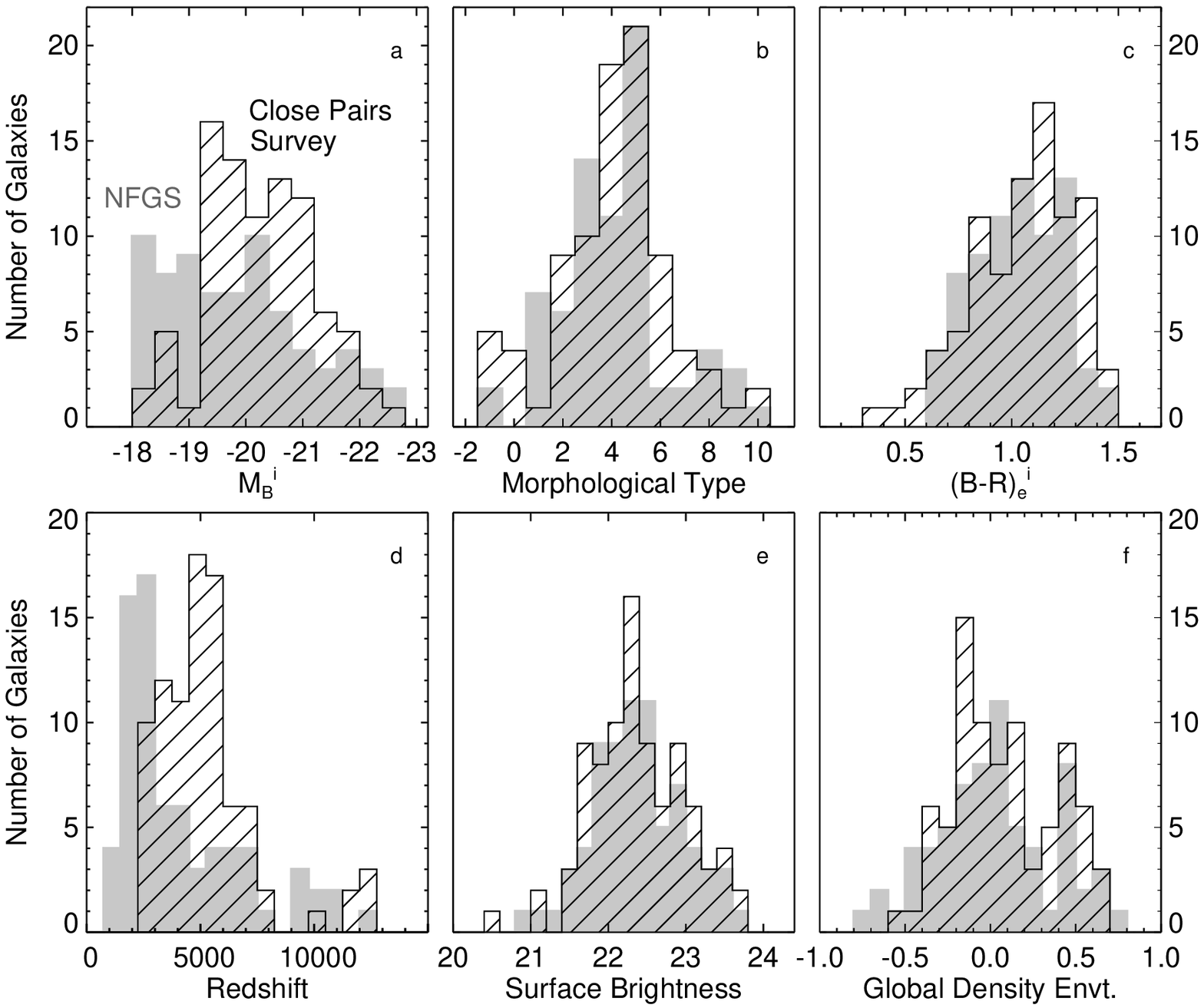}
\begin{figure}[t]
\caption{Property distributions for the 73 NFGS and 88 Close Pairs galaxies
used in our primary TF analysis (emission-line galaxies with $i>40$ and
M$_{\rm B}^i < -18$).  The gray shaded and black cross-hatch histograms
show the NFGS and Close Pairs Survey respectively.  (a) B-band magnitudes.
We have recomputed the Galactic extinction corrections for both surveys
following \citet{schlegel.finkbeiner.ea:maps}.  Internal extinctions are
computed based on \citet{tully.pierce.ea:global} as described in K02,
except with no special treatment of S0 galaxies. (b) Morphological type.
Numbers indicate a modified de Vaucoleurs type system, where -1 = S0, 0 =
S0/a, 1-7 = Sa-Sd, 8 = Sdm, 9 = Sm, and 10 = both Magellanic irregular and
unclassifiably peculiar galaxies.  Classifications for the Close Pairs
Survey were determined by \citet{kannappan.jansen.ea:forming} using the
NFGS as a reference. (c) $B-R$ color within the half-light radius,
corrected for Galactic and internal extinction.  (d) Redshift in $\kms$,
corrected for Virgocentric infall and expressed relative to the Local
Group.  (e) B-band surface brightness at the half-light radius, with no
profile decomposition.  (f) Logarithm of the normalized environmental
density.  Densities are expressed in units of the mean density of galaxies
brighter than M$_{\rm B}\sim-17$ smoothed on 6.7 Mpc scales, using code
adapted from N. Grogin \citep{grogin.geller:lyalpha}.  In these units the
densities of the Virgo and Coma clusters are $\sim$4.9 and $\sim$7.4
respectively, or $\sim$0.7 and $\sim$0.9 in logarithmic units.  The mean
density of 1 (logarithm = 0) represents a field environment.}
\label{fg:twosampprops}
\end{figure}

\begin{figure}[tbh]
\epsscale{0.9}
\plotone{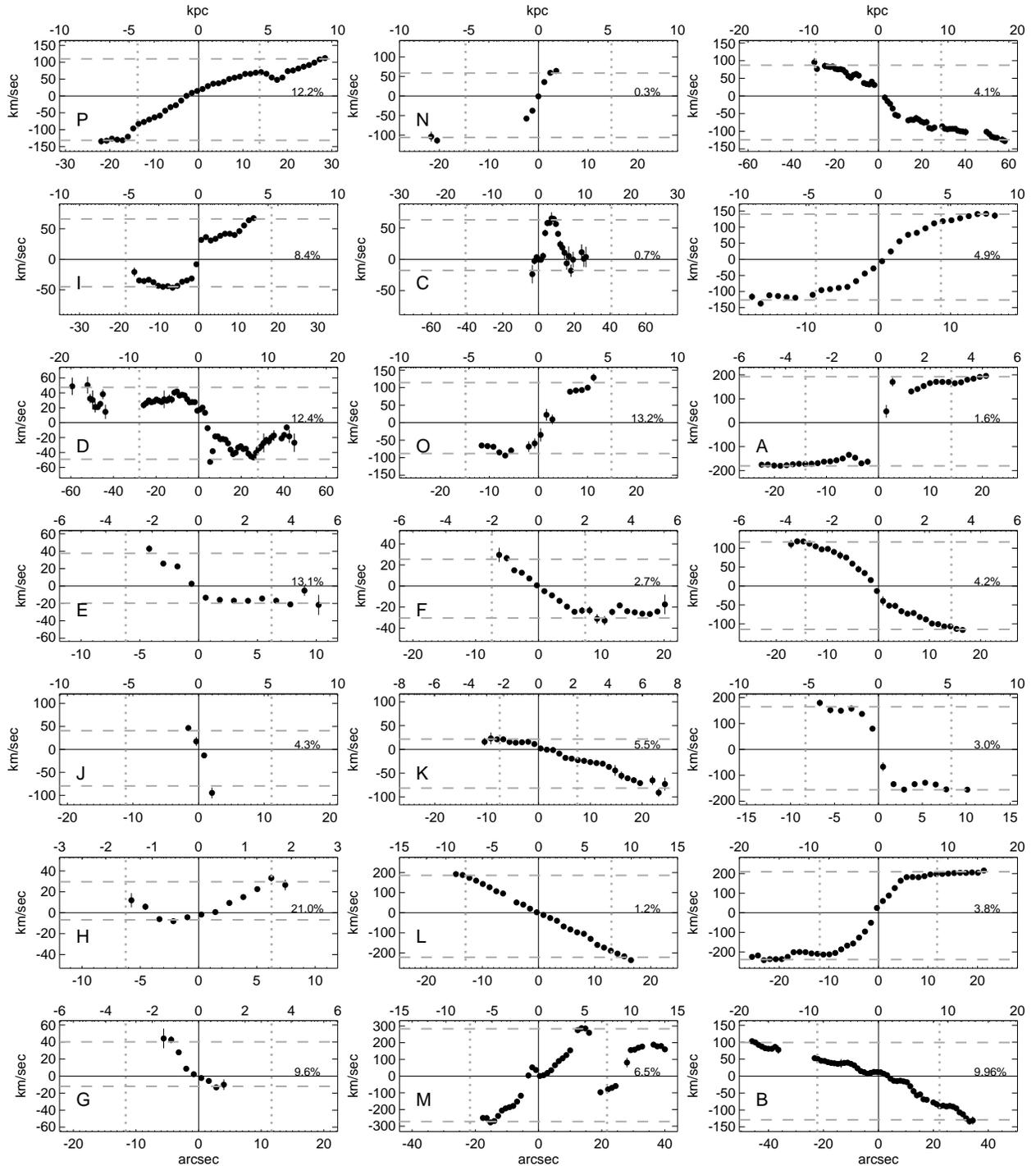}
\caption{Rotation curves from the Close Pairs Survey.  The first column and
the top three examples in the second column are classified as kinematically
anomalous (strongly asymmetric in shape and/or truncated in extent) by our
quantitative criteria.  The remaining examples do not meet these
quantitative criteria, though a few show noticeable abnormalities.  Letter
identifiers correspond to labels used in \S~\ref{sc:pairsresults}.
Measured asymmetries are noted in each panel.  Dotted lines indicate
0.9$r_e$, the reference radius used to evaluate truncation.  Dashed lines
show the maximum and minimum velocities used to define $V_{pmm}$.  Solid
lines mark the coordinate center used for asymmetry calculations in each
galaxy.  We determine this center by minimizing the inner asymmetry (inside
1.3$r_e$) with the spatial center constrained to stay near the continuum
peak and the velocity center allowed to vary freely.  The final asymmetry
is calculated over the full range of radii common to both sides of the
rotation curve.  For example, although the curve for galaxy M appears odd
to the human eye, it shows only modest asymmetry (6.5\%) out to the largest
radius at which flux is present on {\em both} sides of the galaxy.
Truncation is measured from an average of the extent on both sides.}
\label{fg:rcegs}
\end{figure}

\begin{figure}[tbh]
\epsscale{1.}
\plotone{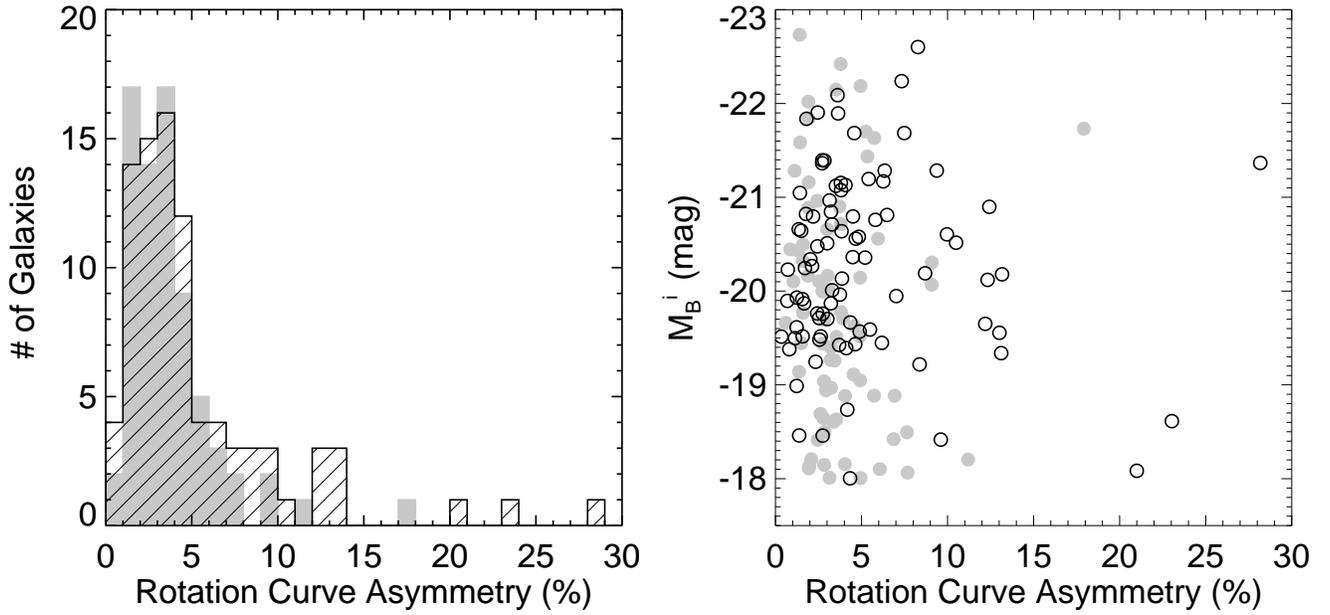}
\caption{(a) Distribution of rotation curve asymmetries for TF sample
galaxies in the Close Pairs Survey (black cross-hatched) and the NFGS (gray
shaded). (b) Luminosity vs.\ rotation curve asymmetry for the same galaxies
(Close Pairs Survey = open black circles; NFGS = gray dots).}
\label{fg:asymdist}
\end{figure}

\begin{figure}[tbh]
\epsscale{0.5}
\plotone{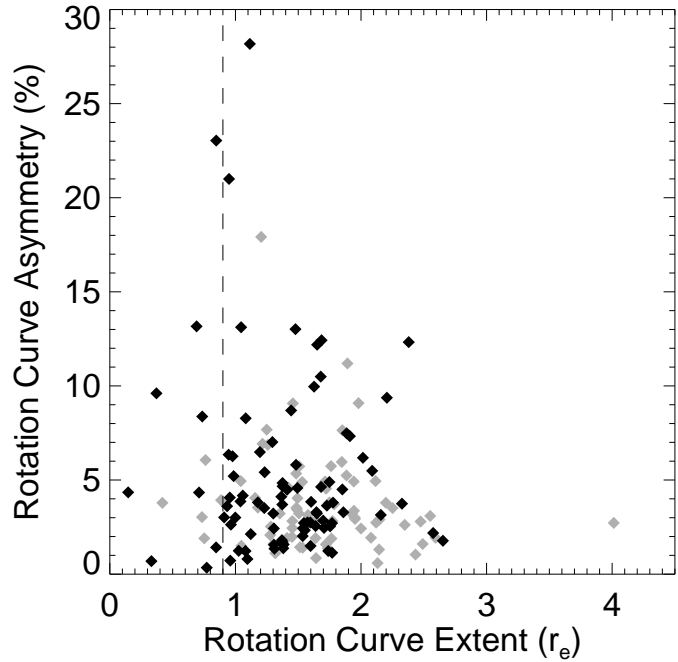}
\caption{Rotation curve asymmetry vs.\ rotation curve extent for TF
sample galaxies from the NFGS (gray) and the Close Pairs Survey
(black). The vertical dashed line marks 0.9$r_e$.}
\label{fg:truncasym}
\end{figure}

\begin{figure}[tbh]
\epsscale{1.}
\plotone{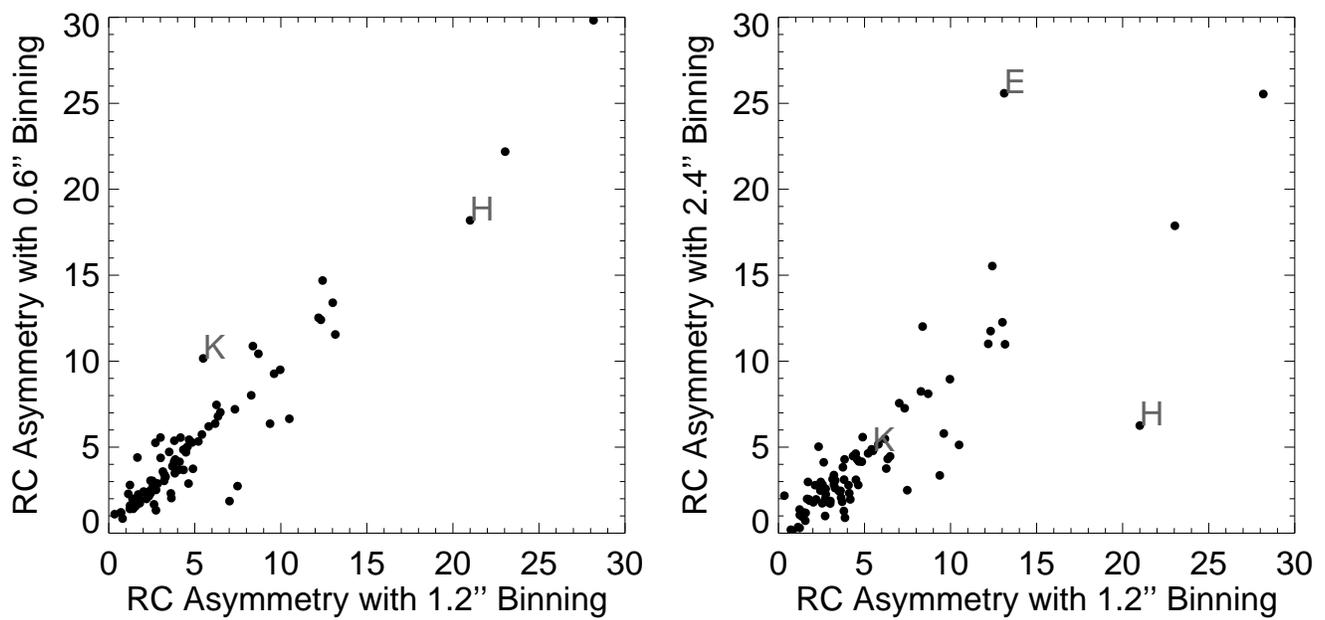}
\caption{Comparison of rotation curve asymmetries measured with
different spatial resolutions, using TF sample galaxies from the Close
Pairs Survey.  Letters refer to specific galaxies discussed in
\S~\ref{sc:pairsresults}.  Galaxy E does not appear in the left panel,
as the lowest resolution data available for this galaxy have
1.2$\arcsec$ binning.}
\label{fg:binning}
\end{figure}

\begin{figure}[tbh]
\epsscale{0.5}
\plotone{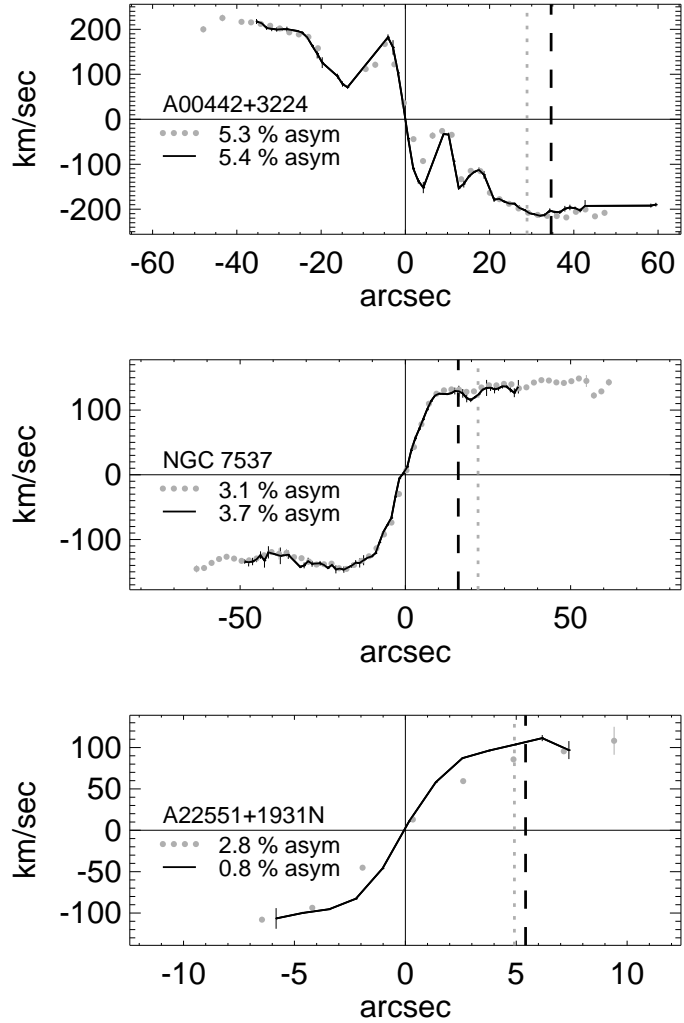}
\caption{Rotation curves for the three galaxies common to both the NFGS
(gray dots) and the Close Pairs Survey (black connected lines).  Asymmetry
measurements from the two surveys correlate well despite small differences
in rotation curve structure caused by the lower spatial resolution of the
NFGS and by different rotation curve extraction techniques (see
note~\ref{fn:struct}).  Vertical lines indicate 0.9$r_e$, the cutoff radius
used to identify truncated rotation curves (dashed = Close Pairs Survey,
dotted = NFGS).  These three galaxies all show adequate rotation curve
extent for TF analysis. However, the variation between surveys illustrates
the potential noisiness of our truncation measure.}
\label{fg:threecmp}
\end{figure}

\begin{figure}[tbh]
\epsscale{1.}
\plotone{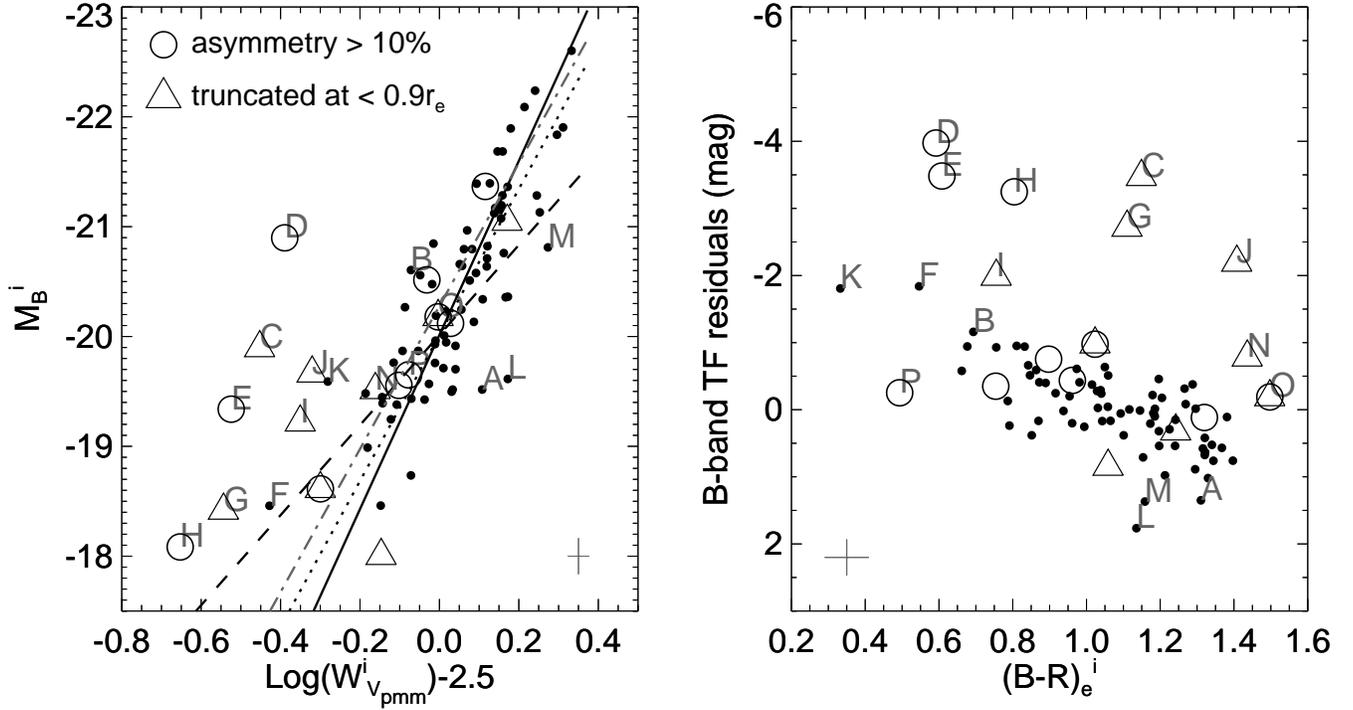}
\caption{TF and CTFR relations for the Close Pairs Survey.  Triangles and
circles mark galaxies with truncated and asymmetric rotation curves,
respectively; these galaxies are omitted from all TF fits except the
dot-dashed line. The solid line shows the inverse-fit TF relation relative
to which TF residuals for the CTFR relation are computed.  The dotted line
shows the bias-corrected forward-fit TF relation, shifted to the zero point
of the inverse-fit relation.  The dashed line indicates how much shallower
the TF relation would need to be to make the CTFR relation statistically
insignificant (see text).  The dot-dashed line shows the inverse-fit TF
relation obtained by restoring kinematically anomalous galaxies to the sample.  Letters refer to
specific galaxies discussed in the text, with letters A--H corresponding to
the labels in B01. Crosses show representative error bars.}
\label{fg:pairsalone}
\end{figure}

\begin{figure}[tbh]
\epsscale{1.}
\plotone{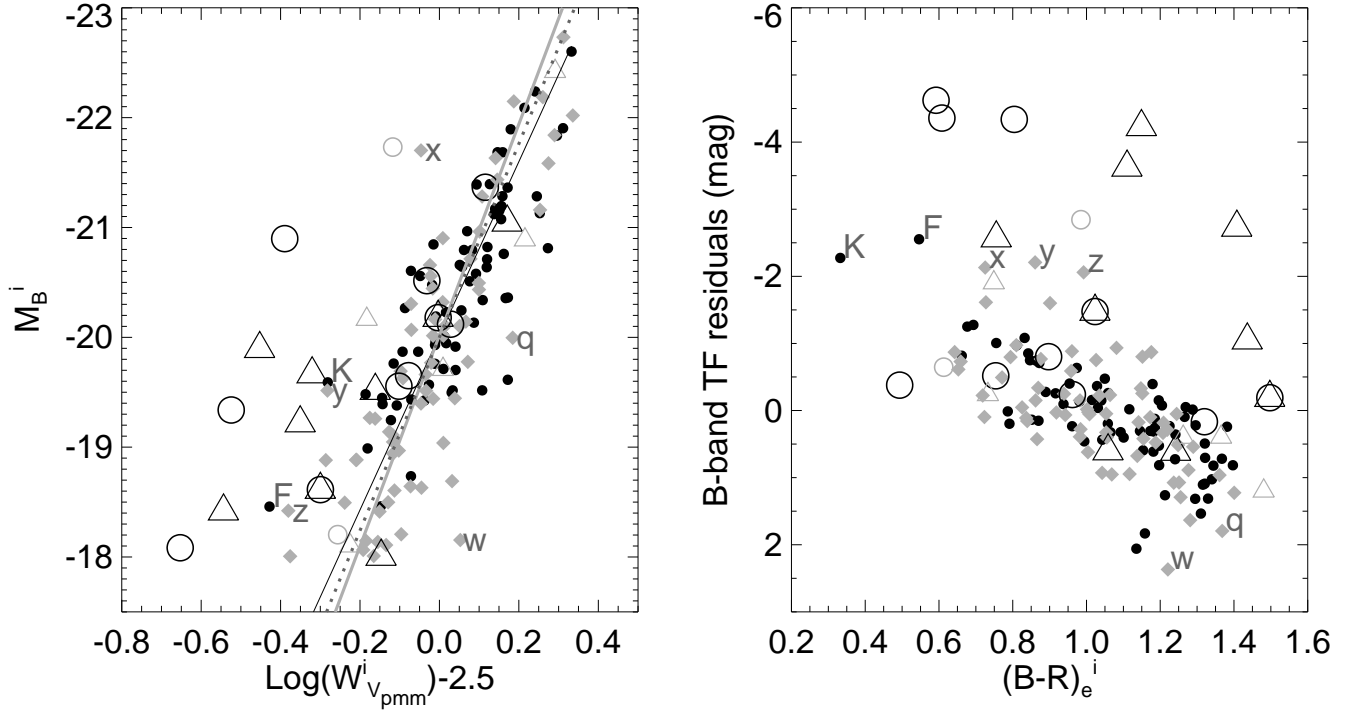}
\caption{TF and CTFR relations for both the NFGS (gray diamonds, small gray
circles and triangles, and solid gray line) and the Close Pairs Survey
(black dots, large black circles and triangles, and solid black line).
Triangles and circles mark galaxies with truncated and asymmetric rotation
curves, respectively; these galaxies are omitted from all TF fits.  The
solid lines show inverse-fit TF relations for the two samples, and the
dotted line shows the intermediate-slope relation obtained by either (a)
omitting galaxies F and K from the Close Pairs Survey or (b) weighting the
NFGS fit in proportion to the Close Pairs Survey luminosity distribution.
We compute TF residuals for the CTFR relation relative to the solid gray
line for both surveys.  Letters indicate galaxies discussed in the text.}
\label{fg:pairsandnfgs}
\end{figure}

\begin{figure}[tbh]
\epsscale{1.}
\plotone{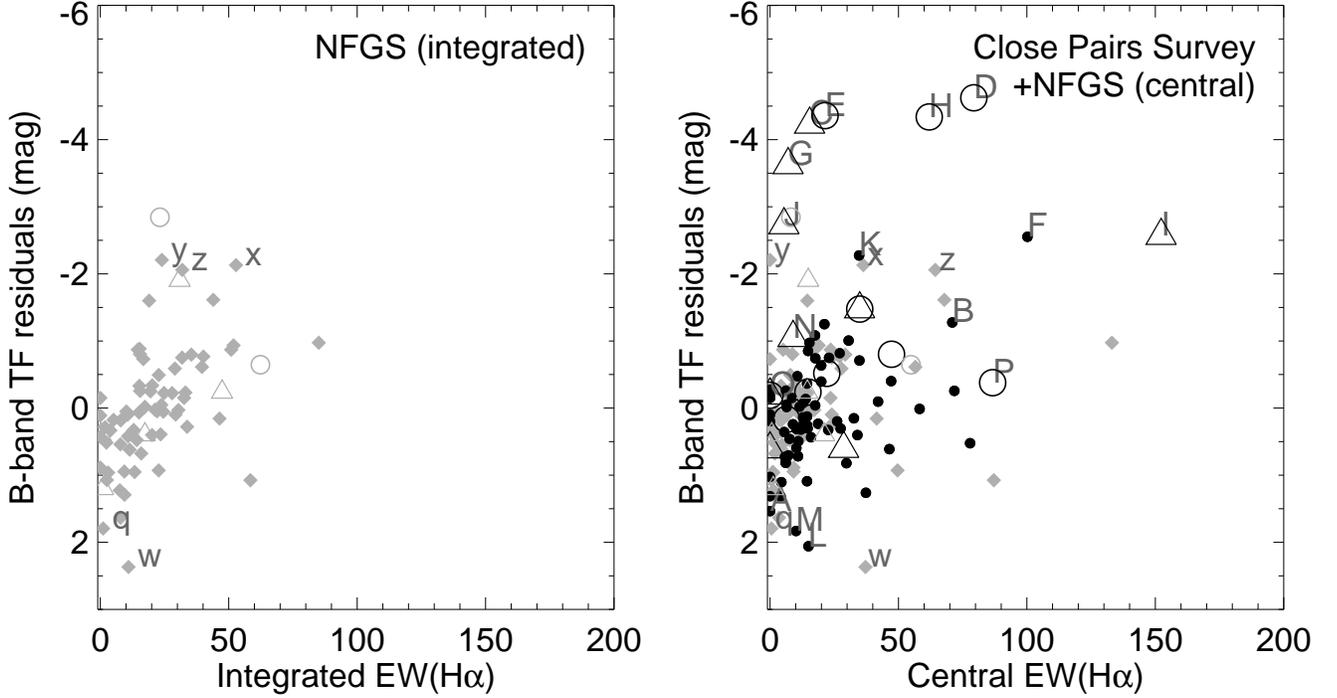}
\caption{TF residuals vs.\ EW(H$\alpha$) for the NFGS (gray diamonds, small
gray circles and triangles) and the Close Pairs Survey (black dots, large
black circles and triangles).  TF residuals are defined as in
Fig.~\ref{fg:pairsandnfgs}.  Equivalent widths in the left panel are
integrated over the entire galaxy
\citep[see][]{jansen.fabricant.ea:spectrophotometry}, while those in the
right panel represent only a central aperture
($\sim$3$\arcsec$$\times$7$\arcsec$ for the NFGS and
$\sim$3$\arcsec$$\times$ a variable length of 2--30$\arcsec$ for the
Close Pairs Survey).  Integrated equivalent widths are not available for
the Close Pairs Survey.  Letters and symbols are as in
Fig.~\ref{fg:pairsandnfgs}.}
\label{fg:bothew}
\end{figure}

\begin{figure}[tbh]
\epsscale{0.5}
\plotone{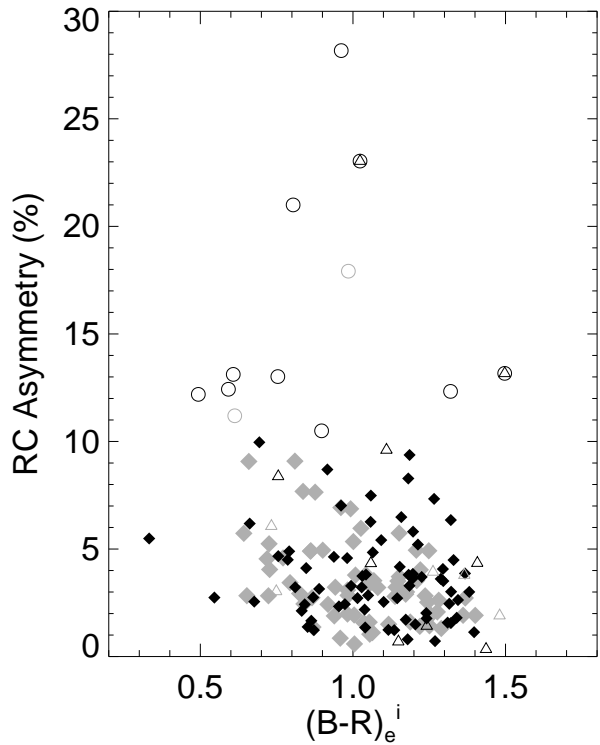}
\caption{Rotation curve asymmetry vs.\ color.  Gray symbols mark NFGS
galaxies and black symbols mark Close Pairs galaxies (diamonds = normal
rotation curves; triangles = truncated rotation curves; open circles =
asymmetric rotation curves). Spearman rank tests give $\sim$3.5$\sigma$
significance for the NFGS correlation and $\sim$2.5$\sigma$ significance
for the Close Pairs correlation, using all galaxies shown.  Excluding
kinematically anomalous galaxies, the correlation strength drops to
3$\sigma$ for the NFGS, while the Close Pairs correlation is no longer
statistically significant.}
\label{fg:colorasym}
\end{figure}

\begin{figure}[tbh]
\epsscale{1.}
\plotone{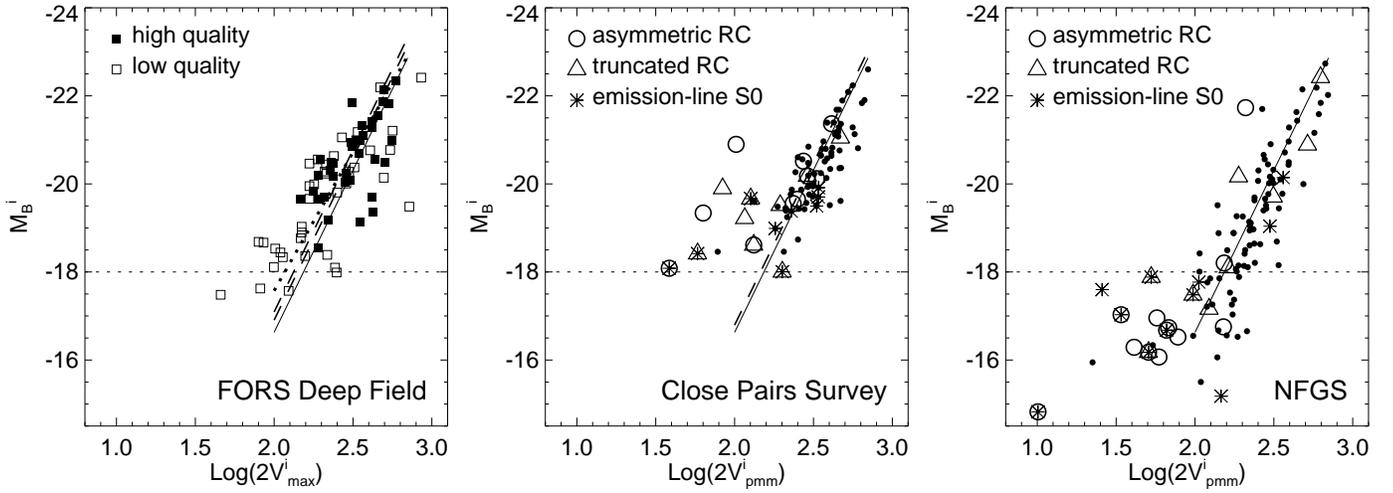}
\caption{Tully-Fisher relations for the FORS Deep Field \citep[general
galaxy population at $z\sim0.1$--1, data courtesy
A. B\"ohm;][]{b-ohm.ziegler.ea:tully-fisher}, the Close Pairs Survey, and
the NFGS (now including dwarf galaxies).  Triangles and circles indicate
severely truncated or asymmetric rotation curves in the two low-$z$
samples.  Likewise, B\"ohm et al.\ flag ``low-quality'' data points in the
high-$z$ sample based on limited radial extent or perturbations in the
rotation curves.  Data points are plotted as filled or open squares
according to their high- or low-quality designations by B\"ohm et al. The
solid line is a fit to the full NFGS Tully-Fisher sample, repeated in all
three panels.  Dashed lines indicate fixed-slope offset fits to the other
two samples, with two lines to show fits with and without the low-quality
data in the FORS Deep Field sample (larger and smaller offsets,
respectively).  The dotted line shows a free-slope inverse fit to the
entire FORS Deep Field sample.  Asterisks indicate emission-line S0
galaxies in the low-$z$ samples.  We have converted the inclinations and
internal extinction corrections for the high-$z$ data to our conventions,
and shifted the data to our cosmology (H$_0=75$, $\Omega_m = 0.3$,
$\Omega_{\Lambda} = 0.7$).}
\label{fg:zieglercmp}
\end{figure}

\end{document}